\documentclass{aastex631}

\pdfoutput=1
\newcommand{\ee}{\end{equation}}

\newcommand{\msun}{{$M_{\odot}$~}}

\newcommand{\lx}{$L_{\rm X}$}

\newcommand{\gtsima}{$\; \buildrel > \over \sim \;$}
\newcommand{\ltsima}{$\; \buildrel < \over \sim \;$}
\newcommand{\prosima}{$\; \buildrel \propto \over \sim \;$}
\newcommand{\gsim}{\lower.5ex\hbox{\gtsima}}
\newcommand{\lsim}{\lower.5ex\hbox{\ltsima}}
\newcommand{\simgt}{\lower.5ex\hbox{\gtsima}}
\newcommand{\simlt}{\lower.5ex\hbox{\ltsima}}
\newcommand{\simpr}{\lower.5ex\hbox{\prosima}}

\newcommand{\cxo}{\textit{Chandra}}

\newcommand{\mstar}{M$_{\star}$}
\newcommand{\reff}{$R\rm{_{eff}}$}
\newcommand{\ks}{$K_{s}$}

\begin{document}

\title{The X-ray Binary-Star Cluster Connection in Late-Type Galaxies}

\author[0000-0002-4669-0209]{Qiana Hunt}
\affiliation{Department of Astronomy, University of Michigan, 1085 S University, Ann Arbor, MI 48109, USA}
\author[0000-0003-0085-4623]{Rupali Chandar}
\affiliation{Ritter Astrophysical Research Center, University of Toledo, Toledo, OH 43606, USA}
\author[0000-0001-5802-6041]{Elena Gallo}
\affiliation{Department of Astronomy, University of Michigan, 1085 S University, Ann Arbor, MI 48109, USA}
\author{Matthew Floyd}
\affiliation{Ritter Astrophysical Research Center, University of Toledo, Toledo, OH 43606, USA}
\author{Thomas J. Maccarone}
\affiliation{Department of Physics \& Astronomy, Texas Tech University, Lubbock, TX 79409, USA}
\author[0000-0002-8528-7340]{David A. Thilker}
\affiliation{Department of Physics and Astronomy, The Johns Hopkins University, Baltimore, MD 21218, USA}
\begin{abstract}
We conduct one of the largest systematic investigations of bright X-ray binaries (XRBs) in both young star clusters and ancient globular clusters (GCs) using a sample of six nearby, star-forming galaxies. Combining complete CXO X-ray source catalogs with optical PHANGS-HST cluster catalogs, we identify a population of 33 XRBs within or near their parent clusters. We find that GCs that host XRBs in spiral galaxies appear to be brighter, more compact, denser, and more massive than the general GC population. However, these XRB hosts do not appear preferentially redder or more metal-rich, pointing to a possible absence of the metallicity-boosted formation of low-mass X-ray binaries (LMXBs) that is observed in the GCs of older galaxies. We also find that a smaller fraction of LMXBs is found in spiral GC systems when compared with those in early-type galaxies: between 8 and 50\%, or an average of 20\% across galaxies in our sample. Although there is a non-negligible probability of a chance superposition between an XRB and an unrelated young cluster, we find that among clusters younger than 10 Myr, which most likely host high-mass XRBs, the fraction of clusters associated with an XRB increases at higher cluster masses and densities. The X-ray luminosity of XRBs appears to increase with the mass of the cluster host for clusters younger than $\sim400$~Myr, while the inverse relation is found for XRBs in GCs.
\end{abstract}



\section{Introduction} \label{sec:intro}

Over the last $\sim$5 years, the detections of merging compact objects by LIGO/VIRGO/KAGRA \citep{ligovirgokagra}, and the discovery of a fast radio burst apparently associated with a globular cluster (GC) in M81 have reignited efforts to self-consistently model the formation and evolution of exotic binary systems  \citep{sridhar21,kirsten22}. A crucial element in these models are X-ray binaries (XRBs) | binaries consisting of compact object and a luminous, typically main sequence donor star. XRBs are a means by which we can observe otherwise invisible compact objects in our galaxy and others. Knowledge of the relative frequency of bright XRBs serves to inform numerical simulations and population synthesis models that aim to understand and predict the formation efficiency of rare and interesting phenomena, such as black hole XRBs, gravitational wave progenitors, fast radio busts, or intermediate-mass black holes \citep[e.g.][]{ozel10,farr11,giersz15,romeroshaw23}. Moreover, XRBs themselves are largely important components to the total X-ray makeup of a galaxy, as they dominate the hard X-ray emissions of a galaxy in the absence of a bright AGN \citep[e.g.][]{fabbiano06,boroson11}.

Stellar clusters in particular are excellent targets in the search for XRBs in external galaxies. Not only are clusters optically brighter than the average extragalactic star, but they are effective factories for XRB formation: they boost the masses of already massive stars, harden already hard binaries, and promote the formation of new binaries through increased dynamical interactions \citep{FabianPringleRees,HillsExchange,Verbunt1987,kundu07,garofali12,ivanova13}. 
Observations of low-mass XRBs (LMXBs; with donor stars of mass $\lesssim$ 1 M$_\odot$) indicate they form hundreds of times more efficiently in GCs than in the field \citep{clark75,katz75,grindlay88,hut92}.  
Over the past 20 years, observations of nearby, massive elliptical galaxies with the Chandra X-ray Observatory and the Hubble Space Telescope (HST) have established that between 20\% and 70\% of bright ($L_X \simgt 5\times 10^{37}$~erg sec$^{-1}$) LMXBs currently reside in GCs \citep[e.g.][]{angelini01,kundu02,jordan04,brassington10,fabbiano10}, and that a near constant $4\%-10\%$ of GCs in ellipticals host bright LMXBs \citep{maccarone03,jordan07,mineo14,luan18}. These high efficiencies suggest that some fraction of field LMXBs may have formed in GCs that subsequently dissolved, or were dynamically ejected \citep[e.g.][]{grindlay84,white02}, though the impact these ejected sources have on the total LMXB population is a subject of debate \citep{piro02,kundu02,kundu07,peacock16,kremer18,lehmer20}. 

Studies also show that, in ellipticals, LMXBs are preferentially found in GCs that are brighter \citep[i.e. more massive; e.g.][]{sivakoff07,brassington10,fabbiano10,riccio22}, more compact \citep[i.e. smaller effective radii and higher densities; e.g.][]{kundu02,pooley03,jordan07,sivakoff07,paolillo11,riccio22}, and redder \citep[i.e. metal-rich; e.g][]{Bellazzini1995,kundu02,kundu03,jordan07,kundu08,sivakoff07,luan18,riccio22}. These surveys have gone a long way towards helping us understand the mechanism by which XRBs form within dense stellar environments.
However, our understanding of the connection between XRBs and star clusters in \textit{spiral galaxies} is comparatively inferior. The complex morphologies, ongoing star formation, and significant dust content of spirals make it extremely challenging to identify clusters in general, let alone to separate ancient GCs | within which LMXBs are produced | from reddened young clusters, which may host higher-mass XRBs \citep{chandar04}. 

It is not obvious whether a comparably high fraction of the LMXB population is expected to be found in GCs of star-forming galaxies. While ellipticals appear to host more GCs per unit halo mass than star forming galaxies \citep{harris15}, the LMXB production efficiency likely decreases at higher redshift \citep{fragos13}.
Tentatively, a smaller fraction of LMXBs are found in GCs across late-type galaxies that have been observed so far | between 2\% and 13\% for extragalactic LMXBs \citep{chandar20, hunt21, hunt22}, though as much as 20\% of the Galactic LMXB population is located within GCs \citep{bahramian22}. The number of these galaxies explored remains too small to definitively prove that this is universally the case, and the depth of recent X-ray observations of these galaxies compared to prior observations of early-type galaxies could play a significant role in these discrepancies.

Furthermore, there is the issue of high-mass X-ray binaries (HMXBs, with donor star mass in excess of 8 M$_\odot$\footnote{The large majority of XRBs with intermediate-mass donors quickly evolve into low-mass XRBs through a short-lived thermal mass transfer phase, so this population is often ignored \citep{podsiadlowski02}.}). Although HMXBs are believed to form in young clusters, most of them are expected to be expelled into the general field population by supernova explosions \citep{kaaret04,sepinsky05,kalogera07,zuo10,poutanen13}. If the natal kick velocities from black holes are smaller than those from neutron stars, then HMXBs with black hole accretors may be more likely to stay within their parent cluster than neutron star XRBs \citep{portegieszwart07}. However, recent work posits the kick velocities of black holes can rival those of neutron stars \citep{repetto17, atri19}. Additionally, tight binary systems may theoretically survive kick velocities well in excess of the cluster escape velocity \citep{sippel13}, and the ejection of black holes from clusters may not be as efficient as previously predicted, as indicated by both more recent simulations and the identification of multiple black holes within compact star clusters \citep{strader12,morscher15}. Therefore, binaries with accreting black holes can be reasonably expected to be found both outside their birth clusters and within them.

Establishing the ratio of HMXBs in clusters versus HMXBs in the field may help illuminate the formation mechanisms and evolutionary histories of compact objects in external spiral galaxies. To date, relatively few papers have looked at XRBs in young clusters, with only a few late-type galaxies examined so far. Current observations suggest between 2\% and 19\% of HMXBs reside within young clusters \citep{rangelov12,chandar20,hunt21,hunt22}. An examination of young cluster populations by \citet{mulia19} find that XRBs in the Antennae galaxies and NGC 4449 prefer more massive and denser clusters among clusters with ages between 10 and 400 Myr, but there does not seem to be a similar preference for HMXBs formed in very young ($<10$ Myr) clusters. Why and whether these patterns hold true for most late-type galaxies remains unknown. \bigskip

In this paper, we undertake a first systematic study of the association between bright XRBs and dense star clusters in a sample of six nearby face-on spiral galaxies that were observed by both the Chandra X-ray Observatory and the HST. We make use of the Physics at High Angular Resolution in Nearby Galaxies (PHANGS)  survey\footnote{\url{https://sites.google.com/view/phangs/home}}, which produced high-fidelity star cluster catalogs \citep[Maschmann et al. in prep.;][]{lee22,thilker22}. We cross-match these catalogs with complete compact X-ray source catalogs based on deep Chandra observations, as compiled by \citet[][hereafter L19]{lehmer19}. The paper is organized as follows: in \S\ref{sec:catalogs}, we describe the X-ray source and optical cluster catalogs, their astrometric alignment, and the positional uncertainties of the observations; in \S\ref{sec:clusterprops}, we describe the cluster properties we analyze and how they are calculated; in \S\ref{sec:xclusters}, we discuss the population of clusters that we determined host XRBs and the properties of those clusters compared to the general cluster population; finally, in \S\ref{sec:stats}, we compare our findings to cluster XRBs found in previous studies, particularly in early-type galaxies. 

\begin{figure}
    \centering
    \includegraphics[width=0.9\textwidth]{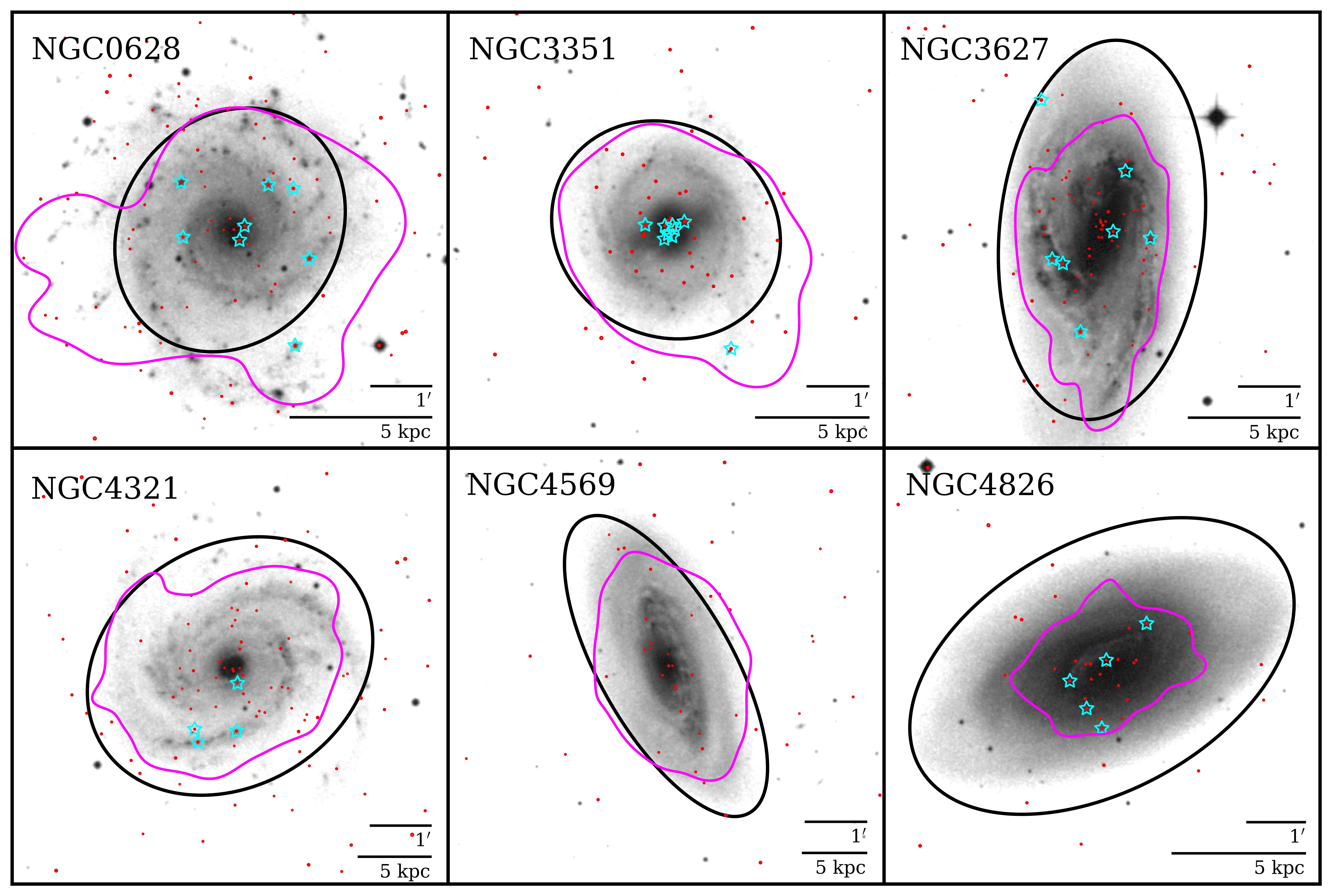}
    \caption{Our galaxy sample, as imaged by the Digitized Sky Survey. Small red circles represent the coordinates and 2$\sigma$ radii of X-ray sources from \citet{lehmer19}, while magenta contours show the outermost density contour of class 1 and 2 clusters in the PHANGS catalog, which we use to denote the PHANGS cluster distribution. Outlined in black are the isophotal ellipses that trace the {K$\rm_{s}~\approx~20$~mag~arcsec$^{-2}$} galactic surface brightness of each galaxy \citep{jarrett03, lehmer19}. Cyan stars represent XRBs that we have identified as having at least one cluster association.}
    \label{fig:galaxies_dss}
\end{figure}

\section{Catalogs} \label{sec:catalogs}

\subsection{X-ray source catalog}\label{subsec:xraycat}
A comprehensive catalog of more than 4000  X-ray sources across 38 nearby galaxies was compiled by L19 from 5.8 Ms of \cxo\ ACIS data. \cxo\ has an excellent angular resolution of $0.\arcsec3$ at its aim-point, which allows us to view nearby XRBs as point-like sources \citep{grimm09}. The galaxies in the sample are all within 30 Mpc and have inclinations~$\lesssim~70^{\circ}$ to our line of sight, which ensures a low degree of local absorption. Because the galaxy selection was based mainly on those covered by both \cxo\ and the \textit{Spitzer} Infrared Nearby Galaxies Survey \citep[SINGS;][]{kennicutt03}, the sample brackets a diverse range of properties, including stellar mass and star formation rate.
L19 also provides a careful assessment of the X-ray completeness for each target galaxy: roughly 3500 X-ray sources are above the $90\%$ X-ray completeness limits of their respective galaxies. 

Additionally, L19 presents a model for estimating the expected contributions from LMXBs, HMXBs and cosmic  X-ray background sources (CXBs) to the total X-ray luminosity function within the footprint of each galaxy. The galactic footprints are defined by the isophotal ellipse tracing the $K_{s}\approx20$ mag arcsec$^{-2}$ galactic surface brightness of each galaxy \citep[L19;][]{jarrett03}. Specifically, L19 provides a functional relationship to estimate the number of LMXBs and HMXBs in each galaxy based on the total stellar mass and star formation rates, respectively, within the isophotal ellipses, as mapped from combined K-band Two Micron All Sky Survey (2MASS) and optical Sloan Digital Sky Survey (SDSS), and combined FUV GALEX and 24$\mu$m \textit{Spitzer} observations. The results of L19's sub-galactic XRB population modeling yield a useful framework for comparing the field vs. cluster XRB populations that we aim to identify with our analysis. 

\subsection{Cluster catalog}
The Physics at High Angular Resolution in Nearby Galaxies (PHANGS)\footnote{\url{https://phangs.stsci.edu}} program is a high-resolution, multi-wavelength survey that utilizes Atacama Large Millimeter/Submillimeter Array (ALMA), HST and Very Large Telescope (VLT) observations to probe spiral galaxies within $\sim20$ Mpc \citep{kreckel19,kreckel21,leroy21,lee22}. Of particular interest to this study is the sample of tens of thousands of compact star clusters identified by the PHANGS-HST\footnote{\url{https://archive.stsci.edu/hlsp/phangs-hst}; \url{https:/archive.stsci.edu/hlsps/phangs-cat/}} survey \citep{deger21,turner21,whitmore21}. These span 38 nearby galaxies with five-band (NUV, U, B, V, and I), high-resolution HST imaging data taken with the WFC3 camera. The sample is purposefully chosen to ensure that the cluster radii are broader than the point-spread function (PSF), thereby minimizing the limitations of instrument resolution. Clusters are identified via a combination of manual and machine-learning classification techniques using a Multiple Concentration Index scheme to select cluster candidates \citep{whitmore21,lee22,thilker22}.  

In this work, we only use centrally concentrated clusters (class 1 and 2), which span all ages. A separate effort by Floyd et al. (in prep) has distinguished ancient GCs from their younger counterparts, based on color selection followed by a careful visual examination. The PHANGS data were obtained from the Mikulski Archive for Space Telescopes (MAST) at the Space Telescope Science Institute and can be accessed via \dataset[DOI: 10.17909/t9-r08f-dq31]{https://doi.org/10.17909/t9-r08f-dq31}. 

    \begin{deluxetable*}{lccccccccc}[t]
\tabletypesize{\scriptsize}
\tablewidth{0pt} 
\tablecaption{Galaxy sample \label{tab:galaxies}}
\tablehead{
\colhead{Galaxy} & \colhead{$\alpha\rm_{J2000}$} & \colhead{$\delta\rm_{J2000}$} & \colhead{Dist.} & \colhead{Scale} & \colhead{log L$\rm{_{X,90\%}}$} & \multicolumn{2}{c}{PHANGS limits$^{b}$} & \multicolumn{2}{c}{Number of sources} \\ [-3ex]
\colhead{} & \colhead{} &
\colhead{} & \colhead{}  & \colhead{} & \colhead{} & \multicolumn{2}{l}{\rule{.8in}{0.01in}}
& \multicolumn{2}{l}{\rule{1.4in}{0.01in}} \\ [-3ex]
\colhead{} & \colhead{} &
\colhead{} & \colhead{(Mpc)}  & \colhead{(pc arcs$^{-1}$)} & \colhead{(erg s$^{-1}$)} & \colhead{(m$_{V}$)} & \colhead{(M$_{V}$)}
& \colhead{X-ray Sources$^{a}$} & \colhead{Clusters$^{b}$} 
} 
\startdata 
NGC 0628 & 03 36 41.8 & $+$15 47 00.5 & 9.84 $\pm$ 0.63 & 35.3 & 36.9 & 24.9  & -5.1 & 155 (47) & 750 (598) \\
NGC 3351 & 10 43 57.7 & $+$11 42 13.0 & 9.96 $\pm$ 0.33 & 45.1 & 37.0 & 25.3 & -4.7 & 65 (39) & 335 (313) \\
NGC 3627 &  11 20 15.0 & $+$12 59 28.6 & 11.32 $\pm$ 0.48 & 45.4 & 37.5 & 24.7 & -5.6 & 105 (64) & 1251 (1236) \\
NGC 4321 & 12 22 54.9 & $+$15 49 20.6 & 15.12 $\pm$ 0.49 & 69.2 & 37.4 & 24.8 & -6.1 & 109 (61) & 868 (866) \\
NGC 4569 & 12 36 49.8 & $+$13 09 46.3 & 15.76 $\pm$ 2.36 & 79.7 & 37.8 & 24.0 & -7.0 & 66 (27) & 540 (527) \\
NGC 4826 & 12 56 43.7 & $+$21 40 57.6 & 4.41 $\pm$ 0.19 & 36.2 & 36.7 & 23.9 & -4.3 & 49 (33) & 175 (175)
\enddata

\tablecomments{Galaxy positions are as reported in L19. We adopt the updated distances used in the PHANGS-HST pipeline \citep{lee22}, and the $90\%$ X-ray luminosity completeness limits (L$\rm_{X,90\%}$) are adjusted from the L19 values to reflect the new distances. The limiting optical magnitude of the PHANGS (V$\rm_{PHANGS,lim.}$) is taken as the optical V-band magnitude of the dimmest cluster identified by PHANGS. The number of X-ray sources and clusters within each galaxy is given, where numbers in parentheses represent the number of sources that fall within the K$_{s}$-band isophotal ellipses, as described in L19. \\
$^{a}$\citet{lehmer19}; $^{b}$\citet{whitmore21,lee22,thilker22}}
\end{deluxetable*}


\subsection{Galaxy Selection}\label{subsec:galaxies}
This paper considers the six spiral galaxies that constitute the intersection between the L19 X-ray source catalogs and the PHANGS-HST cluster catalogs: NGC~628, NGC~3351, NGC~3627, NGC~4321, NGC~4569, and NGC~4826. The target galaxies, which are located between $4-16$ Mpc, are shown in Figure \ref{fig:galaxies_dss}. The L19 X-ray sources are plotted in red, where the radii of each red circle represents the 2$\sigma$ positional uncertainties of the X-ray data. These are calculated, in part, using the $68\%$ and $90\%$ positional uncertainty equations of \citet[][]{kim07}, which incorporates the X-ray counts and off-axis angles of each source provided by L19. The final 2$\sigma$ positional uncertainties include astrometric uncertainties, discussed in \S\ref{subsec:corrections} below, added in quadrature to the positional uncertainties. In addition, the magenta contours represent the outermost distribution of the star clusters within the PHANGS-HST field of view \citep{lee22}. The basic properties of each galaxy, along with the number of X-ray point sources and optically selected clusters, are given in Table \ref{tab:galaxies}. For each galaxy, we utilize the updated distances that were used to identify and estimate cluster properties  \citep[see][]{lee22}, and we adjust the X-ray completeness limit and luminosities of the sample accordingly. The two catalogs capture a total of 549 X-ray sources and 3919 optical clusters across the six galaxies, 271 and 3715 of which fall within the \ks-band isophotal ellipses. 

    \begin{deluxetable*}{lcccccc}[t]
\tabletypesize{\scriptsize}
\tablewidth{0pt} 
\tablecaption{Astrometric corrections\label{tab:astrometrics}}
\tablehead{
\colhead{} & \multicolumn{4}{c}{L19 X-ray sources} &  \multicolumn{2}{c}{PHANGS clusters} \\
\\ [-6ex] 
& \multicolumn{4}{c}{\rule{1.75in}{0.01in}} & \multicolumn{2}{c}{\rule{1.in}{0.01in}} \\ [-1.5ex]
& \colhead{$\Delta_{RA}$} & \colhead{$\Delta_{Dec}$} & \colhead{$\sigma_{RA}$} & \colhead{$\sigma_{Dec}$} & \colhead{$\Delta_{RA}$} & \colhead{$\Delta_{Dec}$} \\ [-2ex]
\colhead{Galaxy} & \colhead{(arcsec)} & \colhead{(arcsec)} & \colhead{(arcsec)} & \colhead{(arcsec)} & \colhead{(arcsec)} & \colhead{(arcsec)}}
\startdata 
NGC 0628 & -0.03 & -0.18 & 0.04 & 0.11 & -0.06 & -0.13 \\
NGC 3351 & -0.21 & 0.29 & 0.48 & 0.45 & 0.01 & -0.01 \\
NGC 3627 & -0.18 & -0.08 & 0.05 & 0.04 & -0.33 & -0.18 \\
NGC 4321 & -0.40 & -0.19 & 0.06 & 0.08 & -0.65 & -0.65 \\
NGC 4569 & 0.23 & -0.07 & 0.06 & 0.05 & 0.01 & -0.02 \\
NGC 4826 & -0.27 & 0.06 & 0.16 & 0.29 & 0.01 & 0.02
\enddata

\tablecomments{The astrometric corrections on the L19 (Chandra) X-ray sources and PHANGS (HST) clusters are given as translational shifts $\Delta$ along right ascension and declination. For the X-ray sources, the standard deviation of the shifts between the L19 reference source coordinates and their optical counterparts on the HST image, $\sigma$, are used to calculate the positional uncertainties of the X-ray sources, which are included in the 2$\sigma$ positional uncertainty calculations.} 
\end{deluxetable*}
 \vspace{-0.75cm}

\subsection{Astrometric corrections}\label{subsec:corrections}

To identify candidate PHANGS-HST cluster counterparts to X-ray sources, we first correct the coordinate offsets between the two catalogs, since observations made by Chandra and HST inherently contain a degree of positional uncertainty that yields an imperfect alignment between the two.
HST images obtained from the Hubble Legacy Archives (HLA\footnote{\url{http://hla.stsci.edu/}}) provide a good reference against which to adjust the coordinates of the L19 X-ray sources, since X-ray catalogs are likely to contain objects with optically visible counterparts such as bright background AGN \citep[e.g., see the methods of][]{hunt21}. The PHANGS-HST catalogs are generated using large mosaics of HST observations, so the catalog positions of PHANGS clusters are shifted slightly when compared to individual HST fields. 

For each galaxy in our sample, we select one HST field from the HLA that was imaged in at least three filters (BVI) and that overlaps with as much of the area enclosed by the $K_{s}$-band ellipses as possible. A handful of sources with visibly identifiable counterparts in the HST field are selected as reference sources for the astrometric calculations of each catalog: background AGN, foreground stars, and isolated GCs in clear proximity to X-ray sources are used for the X-ray catalog, while isolated GCs are used for PHANGS. Ideally, at least 15 reference sources are used per catalog per galaxy, but in some cases (particularly for the L19 catalog), there are fewer than 10 reference sources suitable for use. 

The sources in each catalog are then shifted by the median offsets in right ascension and declination ($\Delta_{\rm{RA}}$~and~$\Delta_{\rm{Dec}}$) between the HST image and the catalog coordinates of the select reference sources. For PHANGS, a simple translational shift is sufficient to properly align cluster coordinates onto the HST image. For the X-ray sources, there is an additional positional uncertainty introduced when combining Chandra and HST observations (e.g. by rotation, instrumental resolution, etc.). These are reflected in the standard deviations of the right ascension and declination correction ($\sigma_{\rm{RA}}$~and~$\sigma_{\rm{Dec}}$) and are factored into the 2$\sigma$ positional uncertainty of each X-ray source as described in \S\ref{subsec:galaxies}. The final astrometric corrections are listed in Table \ref{tab:astrometrics}. After corrections, $95\%$ of the new 2$\sigma$ positional uncertainties of the L19 sources fall between $0.\arcsec21-1.\arcsec96$, with a median of 1.\arcsec07 across all sources. Within the \ks-band ellipses | a region that inherently limits the off-axis angle of each X-ray source and thus the degradation of the Chandra PSF | the maximum 2$\sigma$ uncertainty is $2.\arcsec15$. \bigskip 

\noindent In the following sections, we use the L19 X-ray source catalog as the basis of our investigation. We search for clusters associated with each XRB, where a cluster is determined to be associated with an X-ray source if it falls within a radius equal to the 2$\sigma$ positional uncertainty of said source. We explore the properties of these XRB clusters compared to both the total cluster population and the field XRBs within these spiral galaxies in \S\ref{sec:xclusters}.  We compare our results to previous studies of XRBs within elliptical and spiral galaxies in \S\ref{sec:stats}.

\section{Star Cluster Properties}\label{sec:clusterprops}

The basic properties of the stellar clusters were determined from multi-band HST images of our sample galaxies taken as part of PHANGS-HST \citep[][ Floyd et al. in prep]{turner21,lee22}. The cluster selection methodology is briefly summarized in Section \S\ref{sec:catalogs}; here we describe estimates of the age, mass, and effective radius of each cluster within the galaxy sample. 

\subsection{Age}\label{subsec:clusterage}

The PHANGS-HST survey estimates the age and reddening of each cluster by comparing the measured photometry with predictions from population synthesis models.  Because PHANGS galaxies have formed stars continuously over their lifetimes, they contain clusters spanning a wide age range, from very recently formed clusters with ages $\approx1$~Myr to ancient GCs  with ages of $\approx12$~Gyr. The significant interstellar medium (ISM) in PHANGS galaxies means that breaking the well-known age-reddening degeneracy, where red colors for clusters are either due to  older ages or because they are affected by dust, is critical for correctly age-dating clusters.

Details of the method used to estimate the age and reddening for each cluster are described in \citet{turner21}, and we only provide a brief summary here.  The fluxes measured in five broad-band filters (equivalent to NUV, U, B, V, and I bands) are compared with predictions from the solar-metallicity model for cluster ages between 1~Myr and 12~Gyr \citep{bruzual03}.  Age and reddening are the two free parameters in the fit, and reddening is allowed to vary from an E(B-V) of 0.0 to 1.5~mag.  The software {\em Cigale}  \citep{boquien19} fits the photometry for a cluster to each age-reddening combination in the grid and finds the combination that minimizes $\chi^2$. 

While the PHANGS-HST study estimates ages correctly for many clusters, incorrect determinations have been found to disproportionately affect the ancient GC populations. \citet{whitmore23} and Floyd et al. (in prep) have found that $\sim 80$\% of ancient globular clusters, which comprise $\approx5-10$\% of the PHANGS cluster catalogs, have been incorrectly age-dated, with both too young ages and too high reddening.   
For our sample, we assign an age of 10~Gyr to any GC that is coincident with an XRB, regardless of the age estimated in the PHANGS-HST pipeline. We similarly correct the ages of any intermediate-age ($10-400$~Myr) clusters which have incorrectly been assigned too young of an age from the PHANGS-HST pipeline if a visual inspection indicates they are not heavily affected by dust, using the best-fit age where a low maximum extinction of E(B-V)$<0.1$~mag is allowed. For the rest of the clusters, we adopt the age (and reddening) determined by the default PHANGS-HST pipeline. 

\subsection{Mass}\label{subsec:clustermass}

To estimate cluster mass, PHANGS-HST adopts a mass-to-light ratio predicted by the evolutionary model of solar metallicity clusters by \citet{bruzual03} at the best-fit age obtained in \S\ref{subsec:clusterage}. The mass is then calculated using the extinction-corrected luminosity.

We adopt these default pipeline masses for all clusters where the PHANGS-HST pipeline gives good age estimates, which includes most non-GC clusters. For non-GCs which have incorrect age-dating, a revised mass is calculated by using the M$/$L$_V$ ratio predicted for the new best fit age. To estimate the masses of GCs, we adopt a mass-luminosity ratio of 1.5 \msun/$L_{\odot}$, as given in \citet{chandar07}, where the V-band luminosity of each GC is calculated from the HST V-band magnitude: $L_{V} \equiv 10^{0.4(M_{V, \odot} - M_{V})}$. The absolute V-band Vega magnitude of the Sun ($M_{V, \odot}$) is given by \citet{willmer18} as 4.91 for the WFC3 F555W filter.  The mass estimates are typically uncertain to within a factor of 2 \citep[e.g. see][]{chandar10}. 

\subsection{Effective Radius and Density}\label{subsec:clusterrad}

Floyd et al. (in prep.) measures the effective radius \reff\ of each cluster in the PHANGS-HST catalog using the Ishape routine in BAOlab \citep{larsen99}, a standard software package used to estimate the half-light radii of clusters.
Ishape convolves a series of model clusters having King profiles (with an assumed tidal-to-core radius ratio of 30) with a user-provided PSF. These convolved models are then fitted to the V-band image of each cluster, and the fits are converted to effective radius using the appropriate pixel scale for the WFC3 or ACS camera.  A number of checks were performed to verify the size results, including reproducing the size measurements of clusters published in M51 by \citet{chandar16} and verifying that the Ishape measurements track other size estimates, such as the concentration index (the magnitude difference measured in aperture radii of 1 and 3 pixels).  The distribution of effective (and half-light) radii for ancient globular clusters is consistent with those reported in prior studies \citep[e.g.][]{jordan04b,jordan07,puzia14}.

With the sizes and masses of the clusters, we estimate the density of each cluster by defining the internal half-mass density as $\rho_{h} \equiv 3 M/8 \pi R^{3}_{h}$, where $M$ is the mass of the cluster, and the 3-dimensional half-mass radius $R_h$ is calculated from the 2-dimensional effective radius using $R_{h} = (4/3) R_{\rm eff}$ \citep[e.g. see][]{chandar07, mclaughlin08}. 

    \begin{deluxetable*}{lcccccccccc}[t!]
\tabletypesize{\scriptsize}
\tablewidth{0pt} 
\tablecaption{Properties of XRBs and associated clusters \label{tab:xrb_clusters}}
\tablehead{
\colhead{Galaxy} & \colhead{L19 ID} & \colhead{log L$\rm_{X}$} & \colhead{PHANGS ID} & \colhead{V} & \colhead{U-B} & \colhead{V-I} & \colhead{Age} & \colhead{Mass} & \colhead{R$\rm_{eff}$} & \colhead{Density} \\ [-2ex]
\colhead{} & \colhead{} & \colhead{(erg s$^{-1}$)} & \colhead{} & \colhead{(Mag)} & \colhead{(mag)} & \colhead{(mag)} & \colhead{(log yrs)} & \colhead{(log M$_{\odot}$)} & \colhead{(pc)} & \colhead{(log M$_{\odot}$~pc$^{-3}$)}
}
\startdata 
\multicolumn{11}{c}{Host clusters of singly-associated XRBs} \\
\hline \\[-2ex]
NGC 0628 & L19X038 & 36.8& 481 & -6.30 & -1.38 & 0.51 & 7.28 & 3.47 & 4.80 & 0.13 \\
 & L19X039 & 37.9 & 7257 & -9.75 & 0.20 & 1.22 & 10.00* & 6.04 & 4.65 & 2.74 \\
 & L19X048 & 38.4 & 7375 & -6.60 & -0.01 & 1.75 & 10.00* & 4.78 & 3.38 & 1.90 \\
 & L19X059 & 37.0 & 5558 & -6.21 & -0.67 & 0.43 & 7.88 & 3.75 & 4.27 & 0.56 \\
 & L19X060 & 36.9 & 4602 & -8.13 & 0.49 & 1.57 & 10.00* & 5.39 & 4.16 & 2.24 \\
 & L19X086 & 36.7 & 4841 & -7.94 & -0.01 & 1.27 & 10.00* & 5.31 & 3.38 & 2.43 \\
 & L19X087 & 37.5 & 7497 & -11.30 & | & 0.23 & 6.70 & 5.11 & 4.56 & 1.84 \\
NGC 3351 & L19X014 & 37.1 & 151 & -7.88 & -0.81 & -0.20 & 6.70 & 3.31 & 7.49 & -0.61 \\
 & L19X026 & 38.6 & 3845 & -5.91 & | & 1.50 & 10.00* & 4.5 & 1.37 & 2.79 \\
 & L19X040 & 37.3 & 2881 & -7.85 & 0.24 & 1.26 & 10.00* & 5.28 & 2.20 & 2.96 \\
NGC 3627 & L19X032 & 38.2 & 9731 & -12.32 & -1.10 & 0.49 & 6.00 & 6.17 & 1.95 & 4.00 \\
 & L19X040 & 38.8 & 7367 & -8.04 & -0.05 & 1.13 & 10.00* & 5.36 & 1.26 & 3.75 \\
 & L19X063 & 37.1 & 1421 & -6.78 & -0.50 & 0.99 & 10.00* & 4.85 & 4.28 & 1.66 \\
 & L19X074 & 37.5 & 5530 & -8.65 & -1.04 & 0.62 & 7.51 & 4.41 & 3.20 & 1.60 \\
 & L19X078 & 39.4 & 5876 & -7.51 & 0.44 & 1.20 & 10.00* & 5.15 & 6.94 & 1.32 \\
 & L19X083 & 38.7 & 10673 & -6.69 & 0.39 & 1.01 & 10.00* & 4.81 & 5.03 & 1.41 \\
NGC 4321 & L19X055 & 38.1 & 2466 & -8.21 & 0.13 & 1.14 & 10.00* & 5.42 & 1.70 & 3.43 \\
 & L19X056 & 37.0 & 831 & -9.70 & -0.94 & 0.83 & 6.90 & 4.43 & 1.98 & 2.24 \\
 & L19X070 & 37.1 & 368 & -10.58 & -1.19 & 0.48 & 6.00 & 5.28 & 2.05 & 3.05 \\
 & L19X071 & 37.3 & 878 & -7.88 & -0.77 & 0.72 & 7.66 & 4.13 & 2.12 & 1.85 \\
NGC 4826 & L19X011 & 36.5 & 1885 & -8.31 & 0.07 & 0.76 & 10.00* & 5.46 & 3.16 & 2.67 \\
 & L19X021 & 37.7 & 876 & -8.26 & 0.53 & 1.18 & 10.00* & 5.44 & 1.49 & 3.62 \\
 & L19X023 & 36.8 & 10 & -11.19 & -0.12 & 1.09 & 10.00* & 6.62 & 2.70 & 4.02 \\
 & L19X028 & 36.4 & 33 & -9.98 & -0.18 & 1.09 & 10.00* & 6.13 & 1.49 & 4.31 \\
 & L19X034 & 36.9 & 287 & -9.53 & -0.10 & 1.12 & 10.00* & 5.95 & 1.21 & 4.40 \\
\hline \\ [-3ex]
\multicolumn{11}{c}{Candidate host clusters of multiply-associated XRBs} \\
\hline\\ [-2ex]
NGC 0628 & L19X033 & 36.5 & 3546 & -8.15 & -1.48 & -0.07 & 6.00 & 4.07 & 2.64 & 1.51 \\
 &  & 36.5 & 3557 & -8.12 & -1.20 & 0.17 & 6.00 & 4.33 & 2.21 & 2.00 \\
NGC 3351 & L19X032 & 38.2 & 3382 & -8.12 & -0.99 & 0.08 & 6.70 & 3.57 & 2.71 & 0.97 \\
 &  & 38.2 & 3399 & -9.54 & -0.51 & 1.06 & 8.21** & 5.15 & 4.15 & 2.00 \\
 &  & 38.2 & 3499 & -8.50 & -0.64 & 0.58 & 8.06** & 4.67 & 7.17 & 0.80 \\
 & L19X033 & 38.4 & 3627 & -10.96 & -0.85 & 0.74 & 7.96** & 5.59 & 2.35 & 3.18 \\
 &  & 38.4 & 3660 & -9.90 & -0.98 & 0.46 & 6.60 & 4.5 & 6.08 & 0.85 \\
 &  & 38.4 & 3673 & -9.98 & -1.39 & 0.36 & 6.00 & 4.89 & 6.12 & 1.24 \\
 &  & 38.4 & 3705 & -9.36 & -1.09 & 0.37 & 6.48 & 4.61 & 3.28 & 1.76 \\
 & L19X034 & 37.7 & 3000 & -12.72 & -1.31 & 0.17 & 6.30 & 5.85 & 4.70 & 2.54 \\
 &  & 37.7 & 3009 & -11.92 & -1.16 & 0.36 & 6.48 & 5.6 & 4.60 & 2.31 \\
 &  & 37.7 & 3017 & -10.81 & -1.29 & 0.19 & 6.00 & 5.22 & 6.80 & 1.43 \\
 &  & 37.7 & 3028 & -9.77 & -1.07 & 0.85 & 6.90 & 4.42 & 4.58 & 1.14 \\
 & L19X037 & 38.0 & 3081 & -9.54 & -0.96 & 0.77 & 6.90 & 4.29 & 4.45 & 1.05 \\
 &  & 38.0 & 3159 & -9.27 & -1.07 & 0.67 & 6.90 & 4.11 & 13.09 & -0.54 \\
 & L19X039 & 38.0 & 3624 & -11.60 & -1.35 & 0.23 & 6.00 & 5.49 & 3.03 & 2.75 \\
 &  & 38.0 & 3631 & -9.40 & -1.13 & 0.15 & 6.60 & 4.14 & 1.72 & 2.14 \\
 &  & 38.0 & 3650 & -10.36 & -1.31 & 0.27 & 6.00 & 5.0 & 4.72 & 1.68 \\
 & L19X044 & 37.1 & 3630 & -7.41 & 0.43 & 1.33 & 10.00* & 5.11 & 1.37 & 3.40 \\
 &  & 37.1 & 3712 & -8.53 & 0.01 & 1.15 & 10.00* & 5.55 & 1.37 & 3.84 \\
NGC 3627 & L19X020 & 37.2 & 7021 & -9.01 & -1.31 & 0.12 & 6.48 & 4.33 & 2.62 & 1.78 \\
 &  & 37.2 & 7031 & -9.19 & -1.00 & 0.71 & 6.90 & 4.26 & 3.88 & 1.19 
\enddata
\tablecomments{Singly-associated XRBs refer to those with a single cluster association, while multiply-associated XRBs are those with multiple clusters within 2$\sigma$. The U-B and V-I colors are calculated given the magnitudes reported in the PHANGS, while age, mass, effective radius, and density are obtained as described in \S\ref{sec:clusterprops}. \\
\text{*} XRBs associated with GCs are assigned an age of 10 log yrs due to complications in estimating the ages of GCs (see \S\ref{subsec:clusterage}). \\ 
\text{*}\text{*} Some intermediate-age clusters are assigned manually-corrected ages, as the PHANGS-HST pipeline provides inaccurate age estimates for certain young clusters.}
\end{deluxetable*}


\section{X-ray binary Host Clusters  \label{sec:xclusters}}

Within the fields of view captured by the PHANGS cluster catalog (see Figure \ref{fig:galaxies_dss}, magenta contours), there are 33 X-ray point sources that are spatially coincident with at least one cluster. Depending on the galaxy distance and physical size, a non-negligible fraction of the X-ray point source population may be background AGN, even within the \ks-band ellipses (see Figure 10 in L19). The contamination by CXBs to the total X-ray emissions of the galaxy is estimated by L19; it is likely that unseen CXBs are embedded within the field X-ray source population, which are described and corrected in \S\ref{subsec:cluster_v_field}. 

Observation-based simulations show that the likelihood of a randomly-distributed X-ray source coinciding with an unassociated cluster is small \citep{rangelov12,hunt22}. We further demonstrate the unlikelihood of chance superpositions by performing a similar simulation: for each galaxy in our sample, we randomly scatter $N$ simulated X-ray sources, where $N$ is equivalent to the number of XRBs expected to fall within some area based on the enclosed SFR and \mstar\ (see \S\ref{subsec:xraycat}, \S\ref{subsec:cluster_v_field} and L19 for details). The chosen area in each galaxy is the region covered by both the \ks-band isophotal ellipse and the PHANGS cluster distribution (the intersection of the black ellipses and magenta contours in Figure~\ref{fig:galaxies_dss}). Each simulated source is restricted to this area and assigned a positional uncertainty equal to the median 2$\sigma$ of the L19 sources above the 90\% X-ray completeness limit and within the region. The sources are scattered 1000 times per galaxy, yielding a statistical measure of the number of chance superpositions with a cluster. We find that across all galaxies, the mean and median number of chance superpositions between an XRB and cluster per simulation run is approximately zero, except in the case of NGC 3351, which averaged a single superposition per simulation run (equivalent to 5\% of the simulated XRB population). Thus, we conclude that the probability of a randomly distributed X-ray source intersecting with a cluster is negligible, consistent with previous estimates \citep{rangelov12,riccio22,hunt22}.

However, this might not necessarily be the case for non-random distributions. In a secondary test, we find that shifting the positions of our real X-ray sources by 4$\arcsec$ (a separation sufficiently greater than the 2$\sigma$ positional uncertainty of any source but smaller than the typical distance over which the properties of the target galaxies change significantly) in any direction yields a fairly high probability of the X-ray sources coinciding with at least one cluster: the number of XRB-cluster associations when shifting in any direction decreases by an average of 57\%, and in some cases (e.g. NGC~4321) there is no change. The chance superpositions are mostly driven by younger clusters, largely due to the dense distribution of these sources and HMXBs within active star-forming regions, such as the spiral arms of their host galaxy. Because LMXBs and GCs tend to follow the underlying, radial stellar mass distributions and are generally more isolated and more uniformly distributed than  higher-mass XRBs and younger clusters, we consider any X-ray source that has a GC within 2$\sigma$ to be truly associated with a cluster, while we approach the population of XRBs associated with clusters younger that 400 Myr with more caution. This approach to the GC population is independently supported by prior works \citep[e.g. see][]{riccio22}.

XRBs associated with GCs are almost certainly LMXBs, since only low-mass stars remain in these ancient stellar systems.  The nature of the donor star is less certain for clusters younger than a few hundred million years.  Because the most massive stars tend to sink towards the centers of clusters due to dynamical friction and have been found to be the most active in forming binaries \citep[e.g.][]{trenti06,kaczmarek11,wu21}, very young clusters ($<~10$~Myr) are quite likely to host HMXBs. Clusters with ages between 10 and a few 100~Myr contain both intermediate and low-mass stars. The masses of XRB donor stars in these intermediate-age clusters are currently not well understood.  For neutron star accretors (i.e., the majority of the population), intermediate-mass donors quickly evolve into low-mass stars through a short-lived thermal mass transfer phase \citep{podsiadlowski02,pfahl03,chen16}. 
This likely results in a population of abnormally luminous LMXBs with intermediate-mass donor progenitors. For these reasons, we analyze intermediate-age clusters as a separate population from both GCs and very young clusters. \bigskip

In total, 25 of the 33 point-like X-ray sources are coincident with a single cluster. Of these, 16 are associated with an ancient GC, 5 are associated with a very young cluster, and 4 are associated with an intermediate-age cluster.
Eight additional X-ray sources are coincident with more than one cluster within 2$\sigma$.  This is particularly common in bright, crowded, actively star-forming regions, as stated above. In these cases, it can be challenging to determine the specific cluster host of an XRB. An inspection of X-ray sources with multiple cluster candidates indicates that clusters within 2$\sigma$ of an X-ray source generally have similar ages: 1 X-ray source is associated with 2 GCs, while 7 are each associated with more than 1 cluster younger than $\sim400$ Myr. There are 2 XRBs that are associated with both young \textit{and} intermediate-age clusters: L19X032 and L19X033 in NGC~3351. We also find that clusters associated with the same X-ray source can have drastically different masses, effective radii, and densities. 
Through our analysis, we explicitly indicate which clusters are confirmed XRB hosts (i.e. the only cluster within 2$\sigma$ of a singly-associated XRB) and which are candidate hosts (i.e. one of several clusters within 2$\sigma$ of a multiply-associated XRB). All singly-associated and multiply-associated XRBs and the properties of their (candidate) host clusters are listed in Table \ref{tab:xrb_clusters}. 

One of the galaxies in our sample, NGC~4569, has zero clusters that host XRBs. This may, in part, be due to the fact that NGC~4569 has the brightest X-ray completeness limit of all galaxies in our sample, at 37.8 log L$_{X}$. Indeed, an inspection of X-ray sources with
log L$\rm{_{X}} \geq 37.8$ yields far fewer sources that are spatially coincident with a cluster: only 12 of the 33 XRB clusters have X-ray luminosities brighter than the completeness limit of NGC 4569, with NGC~4826 having none. For this reason, we exclude the clusters from NGC 4569 from the remainder of our analysis, particularly where we calculate statistics using the full star cluster population. 

\begin{figure}
    \centering    \includegraphics[width=0.263\textwidth]{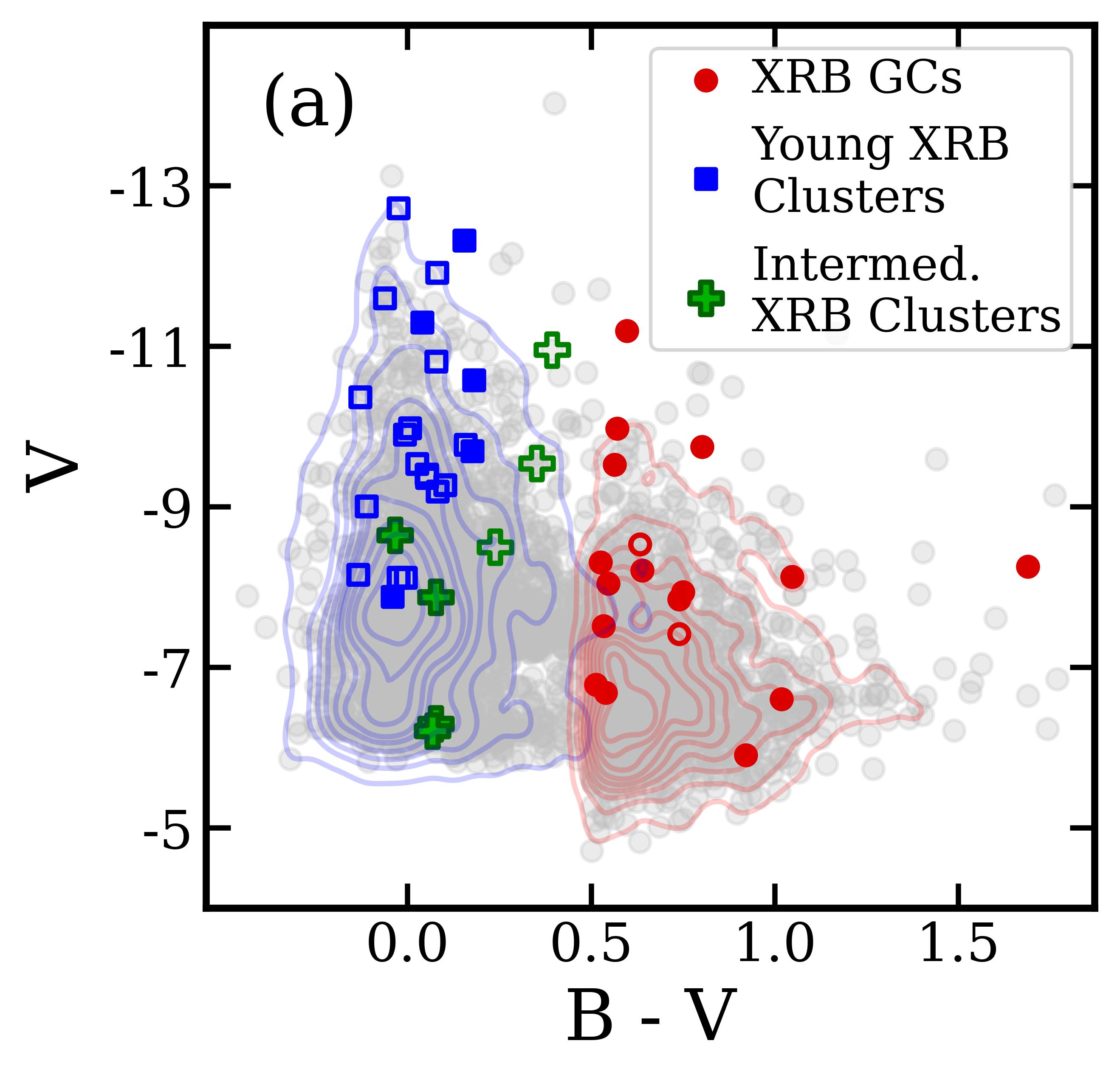}
    \includegraphics[width=0.263\textwidth]{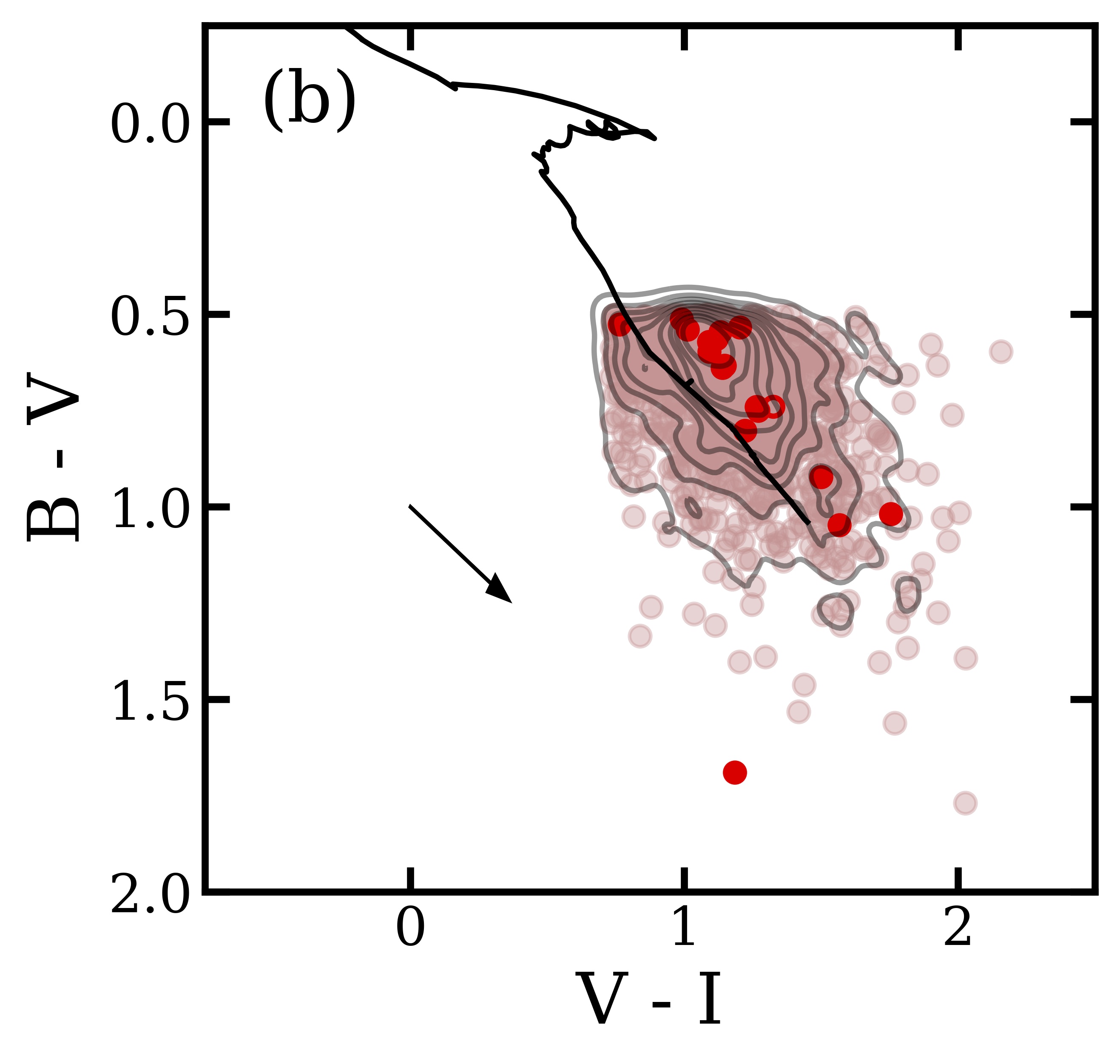}
    \includegraphics[width=0.38\textwidth]{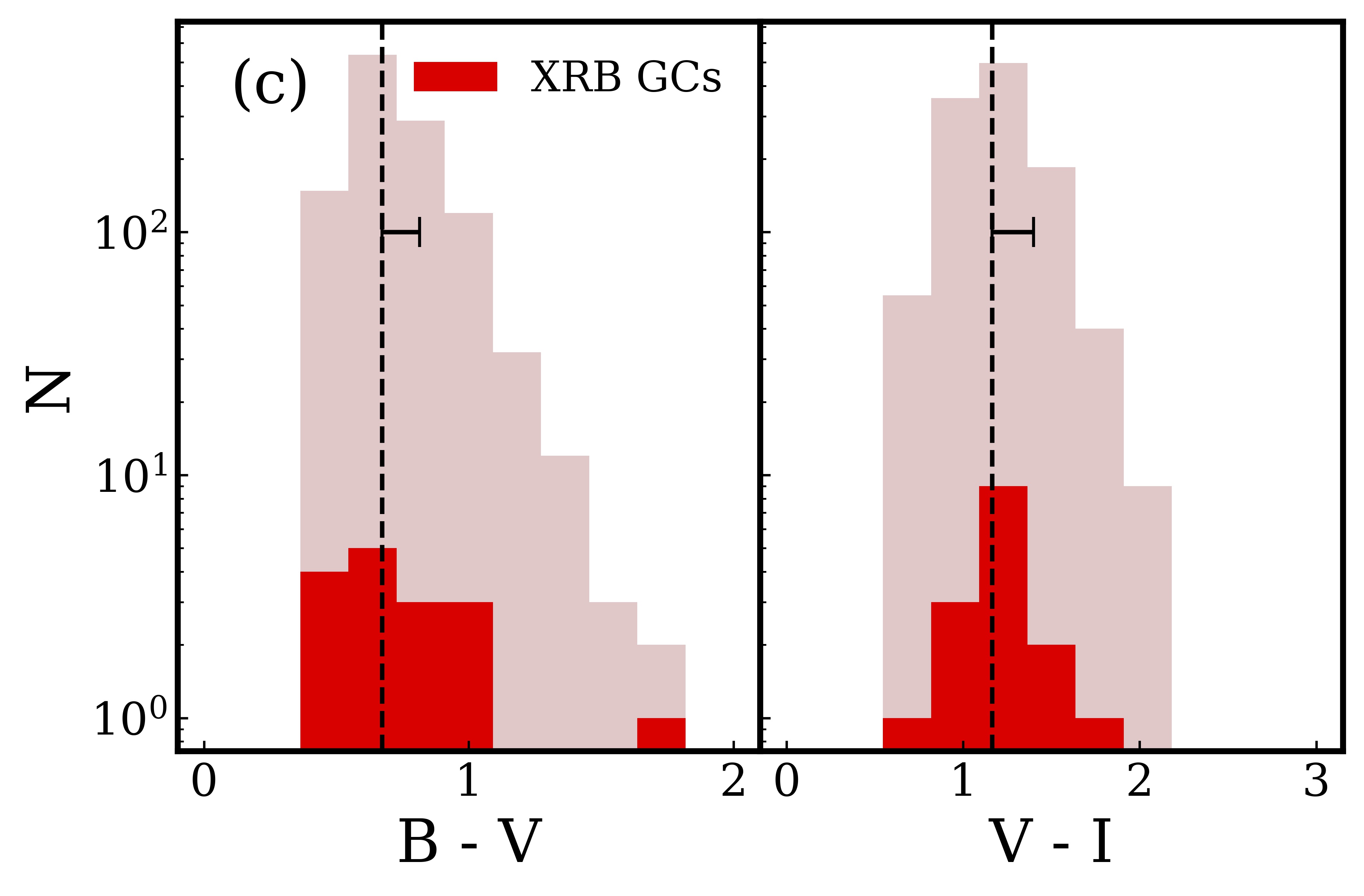}
    \caption{(a) Color-magnitude diagram of all clusters (grey, excluding clusters from NGC 4569) compared to the X-ray emitting GCs, very young clusters, and intermediate-age clusters. The 25 filled symbols represent clusters that are the sole candidate host of an XRB, whereas the 22 open symbols represent clusters that are associated with XRBs with multiple cluster counterparts (1~LMXB and 7 non-GC XRBs). The density contours of the very young cluster and GC populations are shown in blue and red, respectively. (b) Color-color diagram of the GC population, where the evolutionary model for solar-metallicity clusters \citep{bruzual03} is shown in black, and the arrow represents the direction of reddening.
   (c) Histograms of the $B-V$ and $V-I$ colors of GCs. The black dashed line and horizontal bar represent the mean and standard deviation of a Gaussian fit to the total GC population.}
    \label{fig:cluster_cmd}
\end{figure} 

\subsection{Magnitude and color distribution}
The PHANGS cluster catalog includes measurements of cluster magnitudes in 5 HST bands, allowing us to compare the colors and brightness of XRB-host clusters to the general cluster population. Figure \ref{fig:cluster_cmd}a shows the V vs. B-V color-magnitude diagram (CMD) of all clusters, in grey, where the V-band magnitudes were calculated using the apparent magnitudes reported by PHANGS at the galactic distances from \citet{lee22}. The distribution of the GC and young cluster populations are respectively mapped by red and blue contours. 
The CMD shows that XRBs are more common in brighter GCs and very young clusters, compared to the general cluster population in each age group. However, no such preference is apparent among intermediate-age clusters. 

For ancient GCs, we are interested in establishing whether XRBs are more likely to be found in redder (and thus presumably more metal-rich) GCs over blue (metal-poor) ones, as has consistently been found in elliptical galaxies \citep{kundu03,kundu08,sivakoff07,luan18}. Figure \ref{fig:cluster_cmd}b shows the $B-V$ vs. $V-I$ colors of XRB-hosting GCs, compared to the full GC population. For reference, the evolutionary model for solar-metallicity clusters \citep{bruzual03} is plotted in black, and the direction that the clusters may shift due to reddening is indicated by the arrow. To determine whether the XRB GCs are statistically redder than the general GC population, we perform a two-sample independent Wilcoxon test on the $B-V$ and $V-I$ distributions of the two populations using \texttt{ranksum} from the SciPy statistical Python package \citep[see][]{jordan07}. The $B-V$ and $V-I$ distributions are plotted in Figure \ref{fig:cluster_cmd}c. We find that XRB GCs show no color preference in any way that is inconsistent with the underlying GC population. The same test performed on the younger clusters indicates a possible tendency towards bluer colors (and thus presumably younger, more metal-poor systems) for the intermediate-age XRB clusters, although our sample size is too small to determine at this time, and there is a non-negligible chance that at least some of these clusters are spurious associations (as discussed at the beginning of \S\ref{sec:xclusters}).

\subsection{XRB cluster masses}\label{subsec:xrbmass}

\begin{figure}
    \centering
    \includegraphics[width=0.75\textwidth]{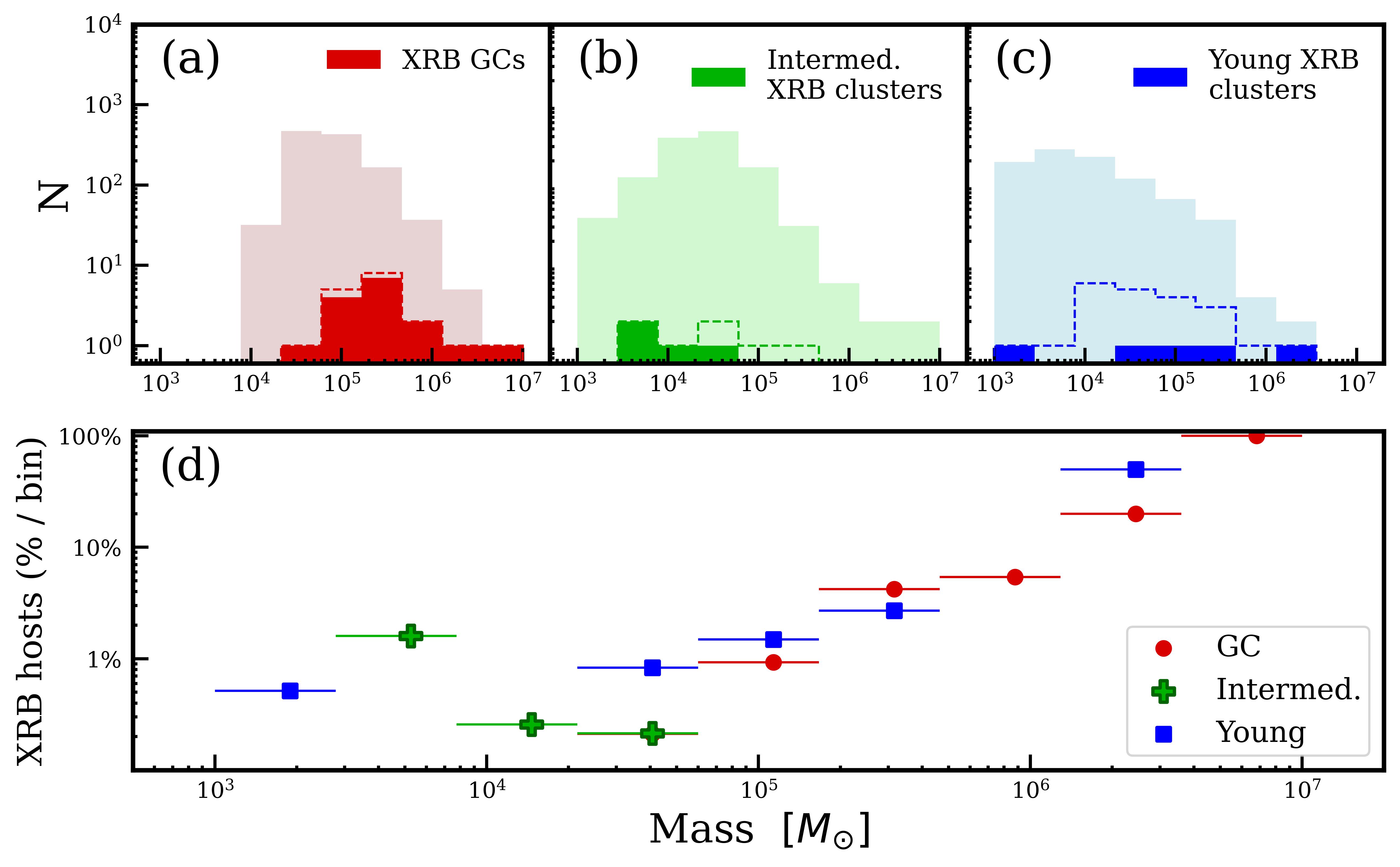}
    \caption{Histogram of cluster masses, separated into (a) GCs, (b) intermediate-age clusters, and (c) young clusters, excluding clusters from NGC 4569. Dashed lines indicate the bins containing the 22 candidate hosts with XRBs that have multiple cluster associations. In panel (d), the fraction of clusters that hosts at least one XRB is given as a percentage of the total clusters per age population per mass bin. The horizontal lines show the boundaries of each bin. 
    }
    \label{fig:clustermasses}
\end{figure}

While the connection between star cluster mass and XRBs has been explored extensively for GCs in elliptical galaxies, little work has been done to date to characterize XRB-hosting clusters in late-type galaxies.
The top panel of Figure \ref{fig:clustermasses} shows histograms of the masses of PHANGS clusters, separated by (a) GC, (b) intermediate-age cluster, and (c) very young cluster populations, where the XRB hosts are highlighted. The bottom panel (Figure \ref{fig:clustermasses}d) illustrates the percentage of clusters that host an XRB as a function of cluster mass. 

It is clear from Figure \ref{fig:clustermasses}a that XRB-hosting clusters occupy the more massive end of the GC mass distribution; that is, bright LMXBs are preferentially found in more massive GCs. In addition, in Figure \ref{fig:clustermasses}d we see that the fraction of GCs that host XRBs increases significantly at higher mass bins. This is true globally and on a galaxy-by-galaxy basis. The fraction of very young clusters that host XRBs also increases with increasing mass, with a nearly 2 order of magnitude difference over the range plotted in Figure \ref{fig:clustermasses}d. The statistics for intermediate-age XRB host clusters are quite poor, and these XRBs do not show any evidence of preferentially forming in higher-mass clusters.

\begin{figure}[t]
    \centering
    \includegraphics[width=0.32\textwidth]{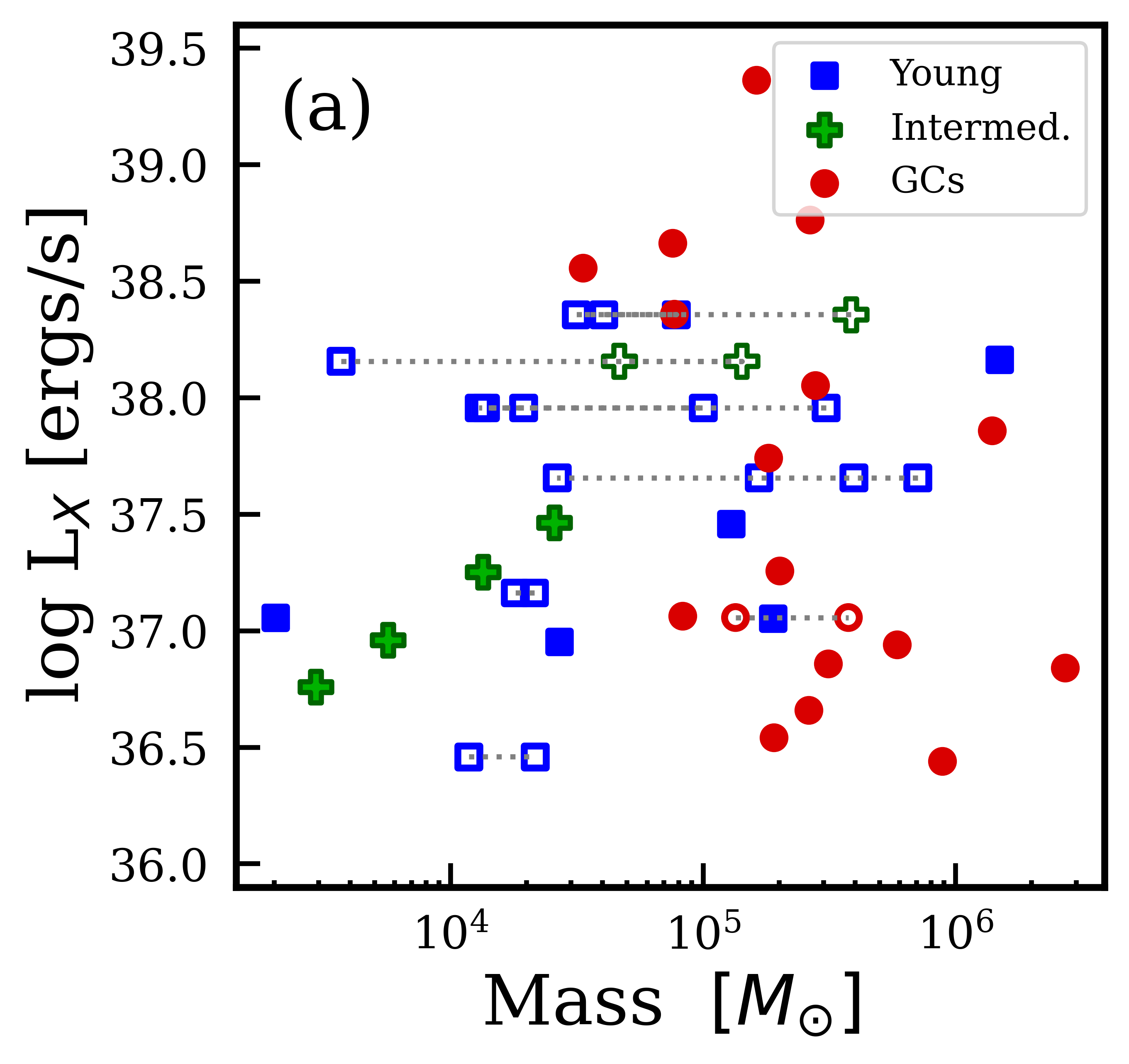}
    \includegraphics[width=0.56\textwidth]{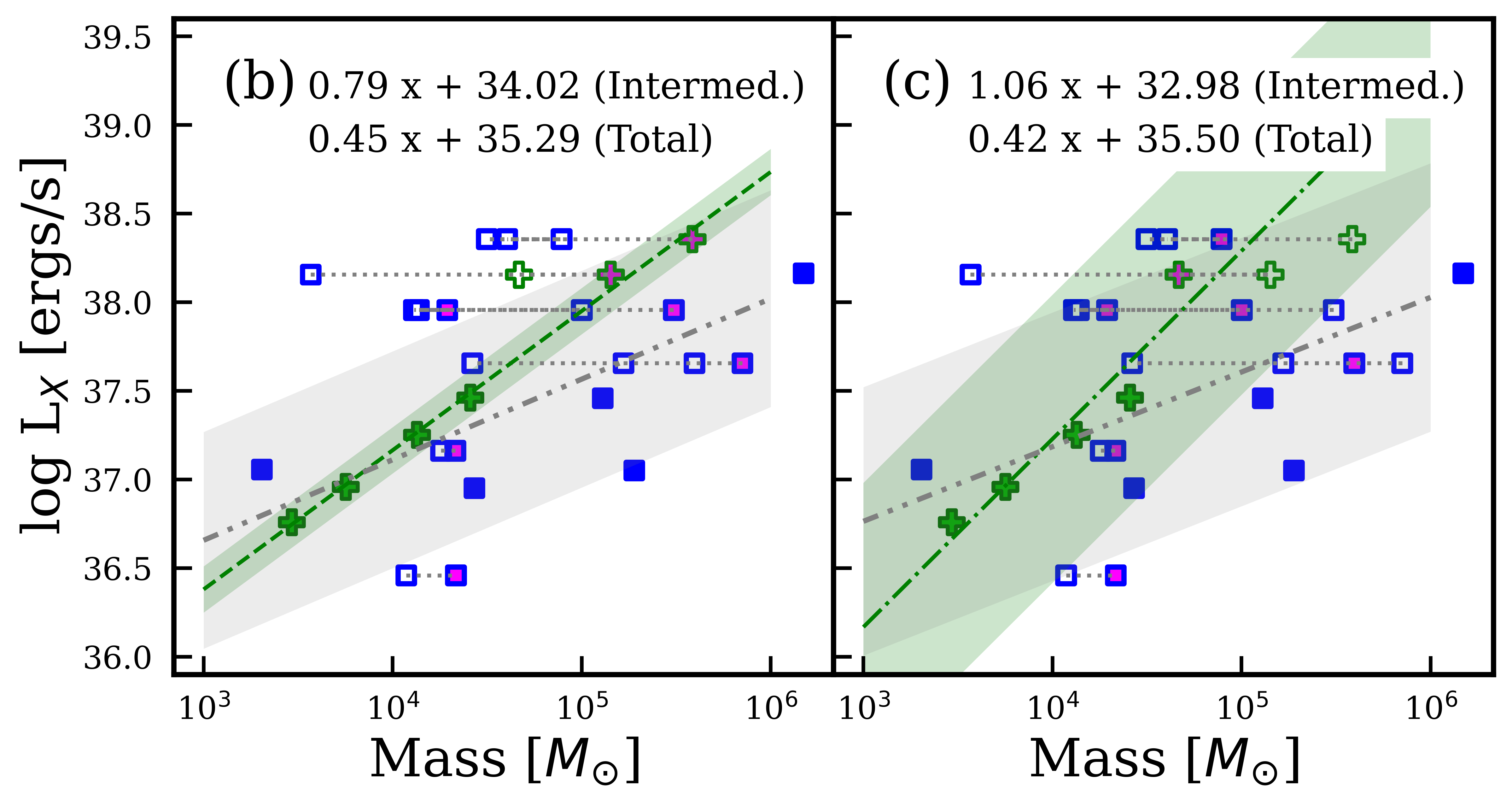}
    \caption{The X-ray luminosities of XRB-hosting clusters as a function of cluster mass. Open symbols represent the 22 candidate clusters of the 8 XRBs with multiple cluster associations. Clusters associated with the same XRB are connected by a dotted line. 
    Panels (b) and (c) represent population sets (i) and (ii) respectively, for which the assumed parent cluster of multiply-associated XRBs are indicated in magenta. The best-fit line for the populations with a statistically significant \mstar-\lx\ correlation | i.e. the intermediate-age XRB cluster and the total non-GC cluster populations | is given (green dashed and grey dash-dotted lines, respectively), where $x = \rm{log}$~\mstar\ in solar units. The linear fits to set (i) for GCs (not shown) is $-0.82x + 42.01$. }
    \label{fig:lx-mass}
\end{figure}

    \begin{table}[t]
\caption{Most significant correlations per population set} 
\centering 
\begin{tabular}{c c c c} 
\hline\hline 
 & Set i & Set ii & Set iii \\ [0.5ex] 
\hline 
Very Young & 0.59$^a$ (0.07) & 0.49$^a$ (0.12) & 0.17$^a$ (0.59)\\ 
Intermediate-age & 0.99$^a$ ($<$ 0.01) & 0.95$^a$ (0.01) & 0.99$^a$ ($<$0.01) \\
Total non-GCs & 0.69$^a$ ($<$ 0.01) & 0.57$^a$ (0.02) & 0.34$^b$ (0.19) \\
GCs & -0.53$^b$ (0.03) & | & -0.49$^b$ (0.05) \\ [1ex] 
\hline 
\end{tabular}
\tablecomments{For each cluster set as described in \S\ref{subsec:m-lx}, the weighted ranks and p-values (in parentheses) for the correlation method that shows the most statistically significant \mstar-\lx\ correlation are shown (excluding results with p-values of 0), where the method from which the statistics are obtained are marked as (a)~Pearson or ~(b)~Spearman. Any rank with an absolute value greater than 0.5 is considered a strong correlation. The sign of the rank represents whether the correlation is positive (increasing) or negative (decreasing). A correlation is statistically significant if it has a p-value $\leq~0.05$.  \label{tab:m-lx_stats}}
\end{table}
 
\subsection{X-ray luminosity vs. cluster mass}\label{subsec:m-lx}

The environment within dense star clusters is conducive to efficient XRB formation, but whether more massive clusters should be expected to retain multiple XRBs at a time is difficult to determine. Observational studies suggest that the brightest metal-rich GCs likely host multiple LMXBs \citep{angelini01,kundu07}; also, multiple LMXBs have been confirmed within Galactic GCs, such as the metal-poor M15 \citep{white01,arnason15}.
\citet{mulia19} approached the question of XRBs in \textit{younger} clusters by comparing the X-ray luminosities to cluster mass, but found no evidence of a correlation for HMXBs. We adopt a similar analysis here.

To quantify any possible correlation between X-ray luminosity and host cluster mass, we run two correlation analyses from the SciPy statistical library: the Pearson correlation, and the Spearman correlation \citep{zwillinger00, scipy}. 
The Pearson correlation measures the degree to which two variables are linearly related, while the Spearman correlation measures for nonlinear, monotonic dependence or association. A correlation is considered statistically significant if it has a p-value~$\leq0.05$, and the strength of the correlation is measured on a scale from 0 (uncorrelated) to 1 (perfect correlation). We analyze the XRB-hosting clusters in each age population (GCs, intermediate-age, and young). We also analyze the combined non-GC population, since the young and intermediate-age populations are spatially intertwined and the physical distinction between them is unclear.

The total cluster sample is shown in Figure \ref{fig:lx-mass}a, including all candidate cluster hosts of multiply-associated XRBs. 
Because 8 of the 33 XRBs have multiple cluster associations spanning a wide range of masses, the results of our correlation analysis depends heavily on the true parent clusters of multiply-associated XRBs. In total, there are 2304 unique combinations of cluster counterparts on which the analyses are run.

Since there is only 1 multiply-associated GC XRB, which has only 2 candidate GC counterparts, the \mstar-\lx\ analysis of the GC population is simple: there are only 2 unique combinations of GC counterparts, and both show, for the first time, a statistically significant \mstar-\lx\ anticorrelation. This is in contrast to the analysis of GCs within M81, which demonstrated the lack of a clear relation \citep{hunt22}, and within NGC 4278, which indicated a weak positive correlation between \lx\ and brighter magnitudes \citep[i.e. higher cluster mass;][]{fabbiano10}. More complicated are the non-GC cluster populations, which include 7 multiply-associated XRBs with 20 cluster candidates. Young and intermediate-age clusters tend to share similar environments, leading to 2 of the multiply-associated XRBs having both young and intermediate-age candidate counterparts. Between young and intermediate-age clusters, there is a total of 1152 unique counterpart combinations. Of these, 62\% yield statistically significant \mstar-\lx\ correlations among the combined (total) young and intermediate-age XRB cluster population. In analyzing these populations independently, we find that \textit{only 6 cluster combinations} (0.5\%)  yield a statistically significant \mstar-\lx\ correlation for young clusters alone. \textit{All} combinations of intermediate-age XRB clusters show a nearly perfect linear \mstar-\lx\ correlation, though it must be noted that there are relatively few intermediate-age XRB clusters, and the 1152 combinations yield only 6 unique intermediate-age configurations. A larger cluster analysis is required to determine the robustness of this trend. 

To better assess how the combination of cluster candidates impacts the correlation models, we highlight the analysis of 3 sets of clusters, for which the parent cluster of each multiply-associated XRB is assumed to be: (i) the most massive candidate cluster; (ii) the median mass candidate cluster, with preference given to the more massive of two clusters if no single median is available; and (iii) the least massive candidate cluster. The results of the correlation analysis on each of these sets is summarized in Table \ref{tab:m-lx_stats}. Interestingly, the trials that include either the highest mass clusters or the median mass clusters (sets i and ii) show  strong positive \mstar-\lx\ correlations at a statistically significant level for both the intermediate-age cluster and the total non-GC populations. The linear fits to the non-GC populations of sets (i) and (ii) are shown in Figure~\ref{fig:lx-mass}b~and~\ref{fig:lx-mass}c, respectively, with the assumed hosts highlighted in magenta. It is clear that, of the three cluster sets, the high-mass trial provides the strongest correlation to the highest statistical significance, followed by the median-mass set. The set which assumes the least massive cluster is the XRB host, set (iii), gives the lowest statistical significance nearly across the board. 

There is no definitive way of knowing which of the 2304 cluster combinations describe the true parent cluster population with the data used in this study. Since we find in \S\ref{subsec:clustermass} that the fraction of clusters that host XRBs appears to increase with increasing mass bins, and because clusters with a greater number of stars (i.e. greater mass) provide more opportunities for the formation of an XRB, it is reasonable to presume that the true host clusters for at least some of the 8 multiply-associated XRBs are likely higher mass. If this is the case, then based on the results from the correlation analyses on set (i) and (ii) in Figure \ref{fig:lx-mass}, there is likely a strong correlation between X-ray luminosity and mass among the total non-GC X-ray clusters population. However, this correlation does not appear to hold among the very young host cluster subsample. For GCs, the \mstar-\lx\ anticorrelation is strongest for the higher-mass candidate host, but is statistically significant either way. 

While, for non-GC XRB hosts, these correlations \textit{may} indicate the presence of multiple XRBs within more massive clusters, there is no indication that these clusters have higher luminosity than the general field XRB population. Furthermore, the correlation for GCs is \textit{negative}, despite being more massive than the non-GC population. Therefore, we conclude that the \mstar-\lx\ correlation for clusters younger than 400 Myr may not be due to the presence of multiple XRBs within X-ray emitting clusters, though the possibility cannot be precluded. This result assumes all XRBs included are indeed found within or near their parent clusters, and the presence of spurious cluster associations may further complicate this relation.

\subsection{XRB cluster radii}\label{subsec:xrbrad}

\begin{figure}
    \centering
    \includegraphics[width=0.75\textwidth]{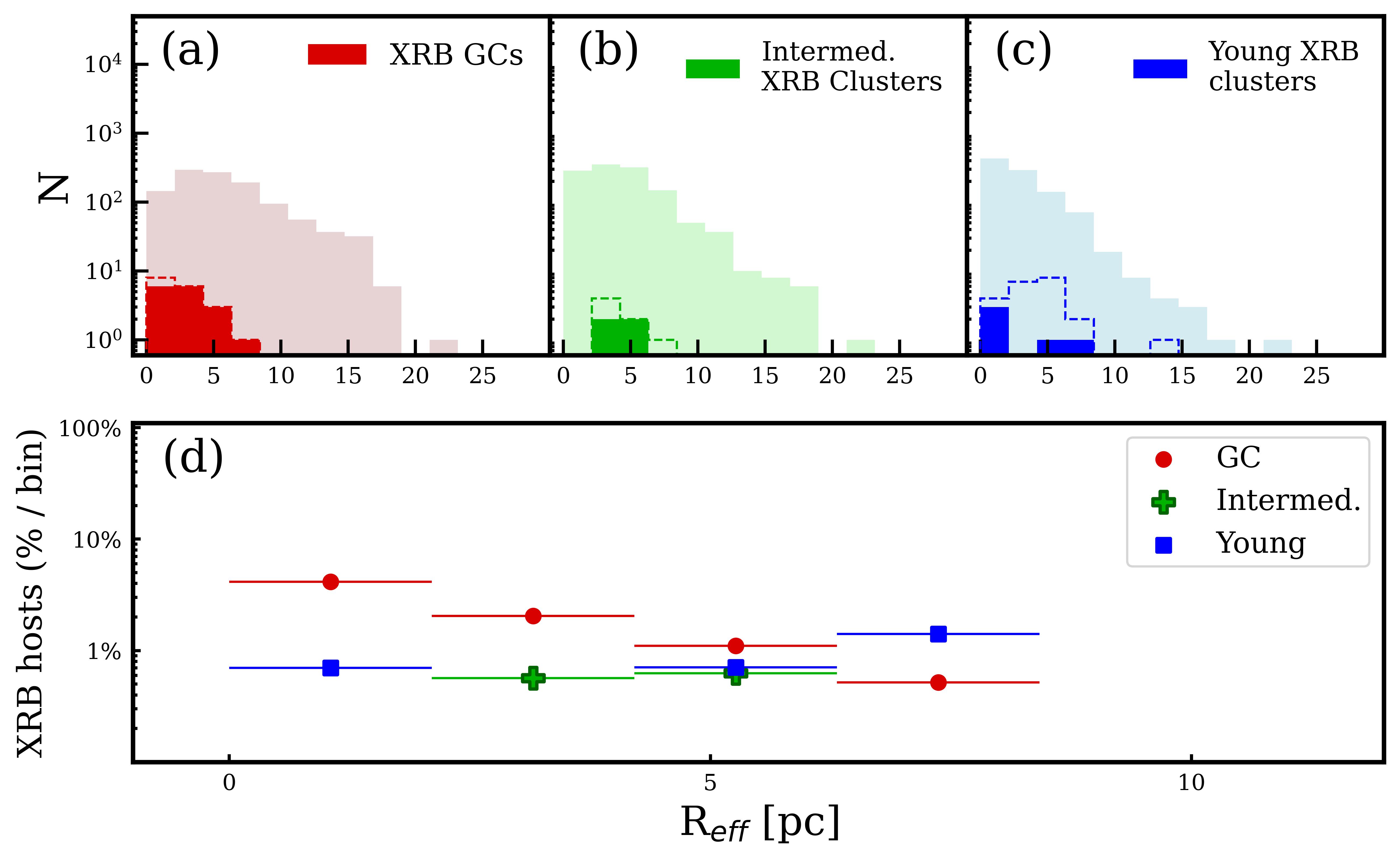}
    \caption{Histogram of cluster effective radii, separated into (a) GCs, (b) intermediate-age clusters, and (c) very young clusters, excluding clusters from NGC 4569. Dashed lines represent candidate hosts for multiply-associated XRBs. In panel (d), the fraction of clusters that hosts at least one XRB is given as a percentage of the total clusters per age population per bin. The horizontal lines show the boundaries of each bin.}
    \label{fig:clusterradii}
\end{figure}

The size distribution of clusters in late-type galaxies remains poorly understood.
It has been proposed that star clusters undergo rapid expansion within their first 10 Myr \citep{bastian12,chandar16,mulia19}. Indeed, there appears to be a preference for XRBs to form in young clusters with smaller radii, a correlation that flattens from intermediate-age clusters to GCs, although this might not necessarily be a consequence of rapid expansion. Generally, young clusters tend to be more compact than older ones \citep{mulia19}. Nevertheless, the size distribution of XRB hosts appears dependent on cluster age. 

Figure \ref{fig:clusterradii} shows the size distribution of PHANGS clusters, binned by radius and divided into (a) GC, (b) intermediate-age cluster, and (c) very young cluster populations. As with Figure 3, panel (d) shows the percentage of clusters that host an XRB per age group per radius bin. In all cases, XRBs are preferentially seen in clusters with smaller radii | although a Wilcoxon test performed on these populations imply only XRB-hosting GCs are statistically smaller than their general cluster population. While the percentage of GCs that host XRBs decreases with increasing radius bin (affirming a strong dependence on cluster size), the same is not seen for intermediate-age and very young clusters. In particular, the fraction of very young clusters that host an XRB appears to \textit{increase}, albeit slightly, with increasing radius bin. If these sources are indeed XRB-hosting clusters and not spurious associations, then this result, in conjunction with the results from \S\ref{subsec:xrbmass}, suggests that cluster mass is more important for XRB formation in young clusters than radius, whereas both cluster mass \textit{and} radius appear to play a strong role in XRB formation in GCs, as seen in early-type galaxies \citep[e.g.][]{paolillo11}.

\subsection{XRB cluster densities}\label{subsec:xrbdens}

\begin{figure}
    \centering
    \includegraphics[width=0.75\textwidth]{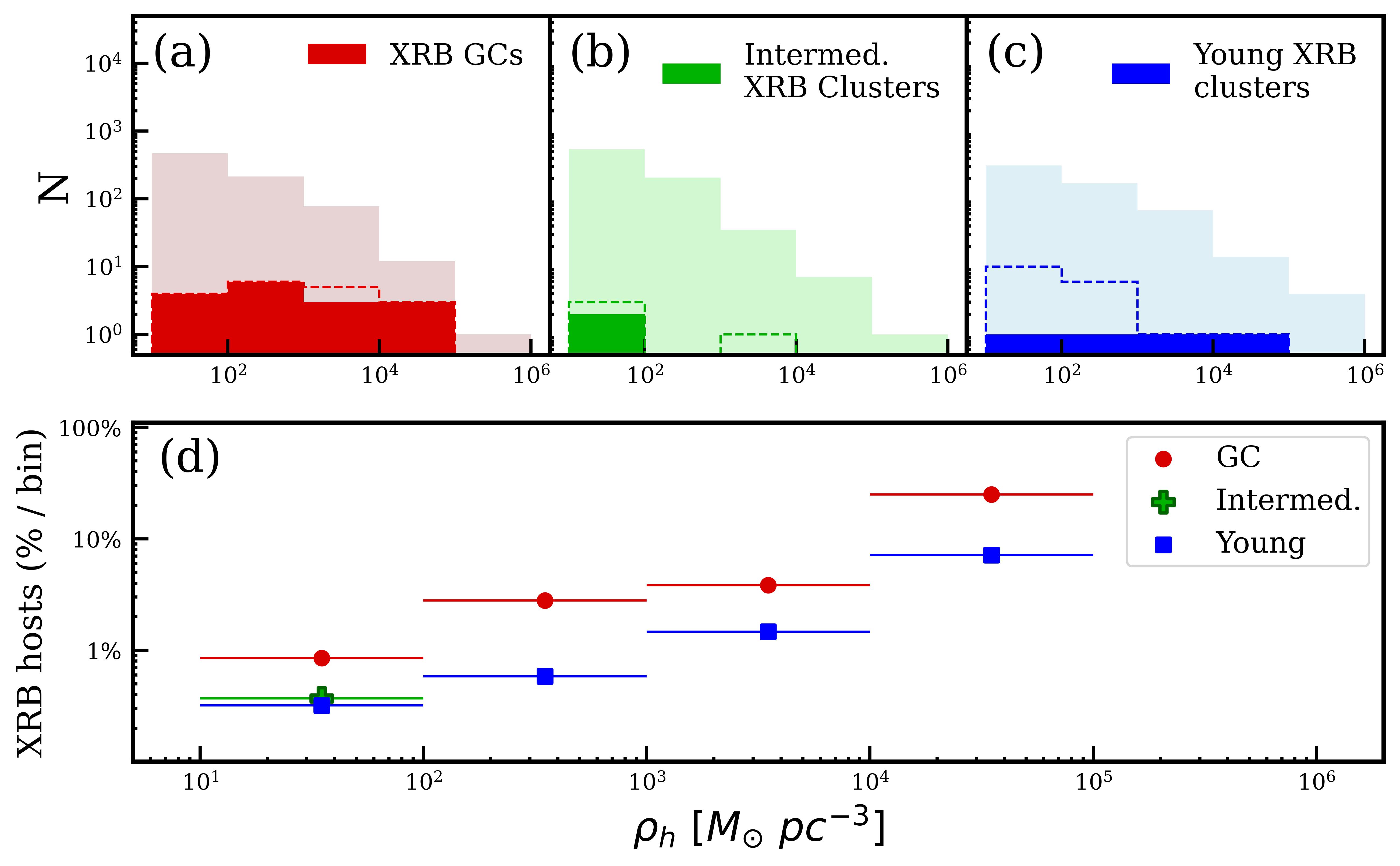}
    \caption{Histogram of cluster densities, separated into (a) GCs, (b) intermediate-age clusters, and (c) very young clusters, excluding clusters from NGC 4569. Dashed lines represent candidate hosts for multiply-associated XRBs. In panel (d), the fraction of clusters that hosts at least one XRB is given as a percentage of the total clusters per age population per bin. The horizontal lines show the boundaries of each bin.}
    \label{fig:clusterdensities}
\end{figure}

Figure \ref{fig:clusterdensities} is a histogram of the densities of (a) GC, (b) intermediate-age cluster, and (c) very young cluster populations, where, as before, panel (d) shows the percentage of clusters that host an XRB per age group per density bin.

The number of XRB-hosting clusters remains relatively flat over increasing density bins for GCs and young clusters, with all of the singly-associated XRBs appearing in the lowest-density bin for intermediate-age clusters (panels a, b, and c).
When observed as a percentage of the total population, however, we see that the fraction of clusters that host XRBs increases drastically with increasing density for both GCs and very young clusters (panel d). The fraction of very young clusters that host XRBs in the highest density bin is a factor of 22 times higher than that of the lowest density bin. For GCs, the difference between the two bins is a factor of 15.

There are a couple of interesting conclusions that can be drawn from Figure \ref{fig:clusterdensities}d. First, the number of GCs and very young clusters that host XRBs is strongly dependent on cluster densities. This may be due to more frequent dynamical interactions in denser clusters,
which could enhance the number of tight binaries that can form in a system \citep{jordan04,peacock09}.
Second, across the total galaxy sample, the fraction of GCs that host XRBs is at least twice that of young clusters for a given density bin. On a galaxy-by-galaxy basis, the difference between age populations is much less dramatic, but is generally consistent with GCs being more likely to host XRBs per density bin. This difference becomes greater if some of the younger clusters are determined to be spurious associations due to the chance superpositon on an XRB with a young cluster in star-forming regions.

\section{XRB-hosting star clusters in late vs. early type galaxies}\label{sec:stats}

The  majority of studies regarding XRBs within star clusters focus on the GCs of elliptical galaxies \citep[e.g.][]{angelini01,jordan04,smits06,peacock16,lehmer20,ferrell21,riccio22}. A few, more recent papers report on XRB associations with GCs in late-type galaxies \citep[e.g.][]{pfahl03,peacock09,generozov18,hailey18,hixenbaugh22}, and fewer still include XRBs in younger clusters \citep[e.g.][]{rangelov12,mulia19,hunt22,avdan22} In this paper, we take the first step towards (i) assessing the fraction of star clusters | both old and young | that host XRBs in late-type galaxies, and (ii) comparing the fraction of field vs. cluster XRBs in late-type vs. early-type galaxies.\\ 

For each galaxy in the sample, the sky area over which the full X-ray source catalog extends is always larger than the sky area encompassing the PHANGS cluster catalog (i.e., the magenta contours in Figure \ref{fig:galaxies_dss}). However, the number of XRBs expected to be observed above the completeness limit, corrected for the CXB, is estimated using models for the XRB X-ray luminosity functions of late-type galaxies generated by L19. These models rely on measurements of \mstar\ and SFR, which are confined to the \ks-band isophotal ellipses. To compare the field-to-cluster XRB ratios of each galaxy, we must restrict our analysis to the region covered by both the PHANGS-HST catalog and the corrected XRB estimates | i.e. the intersection of the isophotal ellipses with the PHANGS contours in Figure \ref{fig:galaxies_dss}. For the sake of consistency, we also use this region for our analysis of XRB clusters vs. the total cluster population.
The \ks-band isophotal ellipses are reasonably well-covered by the cluster catalogs for most of the galaxies; in total, only 2 of the 33 XRB-hosting clusters fall outside of the \ks-band ellipses. Out of the remaining XRB clusters, 22 exceed the X-ray completeness limit of their respective host galaxies.

\begin{table}[t]
\caption{Statistics of clusters inside the \ks-band isophotal ellipse of each galaxy\label{tab:cluster_stats}} 
\centering 
\begin{tabular}{c c c c c} 
\hline\hline 

\\ [-2.5ex]
& & & & \\ [-2ex] 
& GCs (XRB\%) & $\geq10$ Myrs (XRB\%) & $< 10$ Myrs (XRB\%) & Total (XRB\%) \\ [0.5ex] 
\hline 

NGC 0628 & 202 (1.5\%) & 196 (0.5\%) & 200 (0.5\%) & 598 (0.7\%) \\ 
NGC 3351 & 83 (3.6\%) & 79 (0$^{+2.5}$\%) & 151 (0$^{+1.3}$\%) & 313 (1.6\%) \\
NGC 3627 & 536 (0.6\%) & 494 (0.2\%) & 206 (0.5\%) & 1236 (0.4\%) \\
NGC 4321 & 216 (0.5\%) & 327 (0\%) & 323 (0\%) & 866 (1.2\%)\\ 
\textit{NGC 4569} & \textit{110 (0\%)} & \textit{371 (0\%)} & \textit{46 (0\%)} & \textit{527 (0\%)} \\ 
NGC 4826 & 43 (7\%) & 73 (0\%)  & 59 (0\%)& 175 (1.7\%) \\ [1ex]
\textbf{Total*} & 1080 (1.2\%) & 1169 (0.1$^{+0.1}$\%) & 939 (0.5$^{+0.2}$\%) & 3188 (0.7\%) \\ 
$\rm^{(excl.~NGC4569)}$ & & & & \\
\hline 
\end{tabular}
\tablecomments{Percentages represent the fraction of clusters that host an XRB within the PHANGS-HST catalog, above the X-ray completeness limit of each galaxy. Upper limits indicate how the values may change based on the age of the true parent cluster of multiply-associated XRBs. The percent of clusters younger than 400 Myr that host an XRB may be inflated due to the non-negligible likelihood of a chance superposition of an XRB and young cluster in densely population regions (e.g. spiral arms). \\ 
\text{*}The total values exclude clusters from NGC 4569, which may be significantly affected by the high X-ray completeness limit of observations.}
\end{table}

\subsection{Properties of XRB-hosting clusters\label{subsec:xclusters_v_clusters}}

In elliptical galaxies, between $4$ and $10\%$ of GCs host LMXBs \citep[e.g.][]{maccarone03,jordan07,brassington10,mineo14,luan18}.
Analyses of these XRB hosts suggest they are more massive, more luminous, radially smaller, denser, and redder than the general GC population \citep[e.g.][]{kundu02,sivakoff07,peacock09,brassington10,paolillo11,riccio19}. Further investigations suggest that metallicity | more so than cluster age |  plays a role in the production of LMXBs \citep{kundu03,kundu08,sivakoff07,fabbiano06,peacock10,dago14,luan18,riccio22}. 

In late-type galaxies, the percentage of GCs that host LMXBs is still an unsettled matter, though a recent study of M81 suggest a fraction consistent with that of elliptical galaxies | around $4\%$ \citep{hunt22}. Similarly, XRB-hosting GCs in M81 tend to be more compact, denser, and more massive than the rest of the GC population \citep[][see also \citealt{bregman06} in the context of the Milky Way galaxy]{hunt22}. 
As far as young clusters go, a study of XRB-hosting clusters younger than 400 Myr in the Antennae galaxies and NGC 4449 suggests that more massive and denser intermediate-age clusters may be more likely to host XRBs, but that this preference is not reflected in the the population of clusters younger than 10 Myr \citep{mulia19}. \\

The total number of clusters and the percent that host XRBs in each galaxy is given in Table \ref{tab:cluster_stats}. Among these late-type galaxies, the percentage of clusters that host bright XRBs above the completeness limits of their host galaxy varies 
between $0.5-7\%$ for GCs, $0.2-2.5\%$ for intermediate-age clusters, and $0.5-1.3\%$ for young clusters, excluding NGC 4569 (for which no XRB-hosting clusters were identified). Combining all clusters within our galaxy sample (excluding NGC 4569), the percentage of XRB hosts among GCs, intermediate-age clusters, and young clusters is 1.2\%, and 0.1$^{+0.1}$\%, 0.5$^{+0.2}$\%, respectively. Tentatively, a total of 0.7\% of all compact star clusters within the PHANGS catalog, irrespective of age, host XRBs. As a word of caution, we note again that the statistics for clusters younger than 400 Myr may be inflated due to the high probability of a chance superposition with an unrelated cluster in densely populated, active regions, as determined in \S\ref{sec:xclusters}.

XRBs in late-type galaxies appear to prefer smaller, denser, brighter, and more massive GCs, just as they do in elliptical galaxies (see Figures \ref{fig:clustermasses}, \ref{fig:clusterradii}, and \ref{fig:clusterdensities}d). This is consistent with the notion that clusters with more stars (that is, more massive and brighter) that are in closer proximity to each other (denser and smaller) provide ample opportunities for the formation of a close binary \citep{smits06,jordan07,peacock09,peacock10}. 

In contrast to elliptical galaxies, we find that XRB-hosting GCs in the star-forming galaxies under consideration are \textit{not} redder than the general GC population | rather, their colors are consistent with those of the full population (see Figure \ref{fig:cluster_cmd}c). For early-type galaxies, it is believed that metallicity plays a large role in this correlation, possibly due to enhanced tidal capture rates and core concentrations (smaller radii) in higher-metallicity GCs, which in turn facilitates the formation of LMXBs \citep{jordan04b,schulman12,ivanova13,vulic18,hixenbaugh22}. It is interesting to note that, while GCs typically display bimodal colors corresponding to metal-rich and metal-poor populations \citep[e.g.][]{peacock16,hixenbaugh22}, the GCs in our sample lack a clear distinction between red and blue GCs (Figure \ref{fig:cluster_cmd}b and \ref{fig:cluster_cmd}c). This could be due to the fact that (i) the metal poor ``mode" becomes very broad when different late type galaxies (each with different star forming histories) are added together, and (ii) the metal rich mode may be triggered by mergers, which are rarer in spirals \citep{muratov10,pfeffer23}. 

Regarding younger clusters, we find strong evidence that brighter, more massive, and denser clusters younger than 10 Myr are more likely to host XRBs than other clusters of similar ages (see Figures \ref{fig:cluster_cmd}, \ref{fig:clustermasses}d, and \ref{fig:clusterdensities}d). The sample of intermediate-age XRB-hosting clusters (younger than 400 Myr) is too small to draw a solid conclusion, but their properties appear consistent with the broader intermediate-age population. This is possibly in contrast to NGC~4449 and the Antennae galaxies, which show a tentative correlation between XRB formation and mass and density in intermediate-age clusters, but no such correlation in young clusters \citep{mulia19}. Larger statistics are needed to definitively confirm whether a trend exists, as only 11 intermediate-age and 17 young XRB-hosting clusters were observed in the NGC~4449 and the Antennae galaxies, compared to the 17 non-GC XRBs we analyze here. Furthermore, the likelihood of chance superpositions with unrelated clusters causing spurious associations to be included in our analysis may further complicate matters. It is interesting that such strong correlations between the presence of an XRB and cluster mass and density are identified; however, this could point to the associations of singly-associated XRBs, at the very least, being real, and that the background contamination is minimal.

    \begin{table}[t]
\caption{Statistics of X-ray sources inside the isophotal ellipse and PHANGS footprint of each galaxy \label{tab:xrb_cluster_stats}} 
\centering 
\begin{tabular}{c c c c c c c c c} 
\hline\hline 

\\ [-2.5ex]
& & & & \multicolumn{2}{c}{Estimated XRBs} & \multicolumn{3}{c}{Observed XRB-hosting clusters} \\ [-2ex] 
& \mstar$\rm{_{,encl}}$ & SFR$\rm{_{encl}}$ & X-ray & \multicolumn{2}{c}{\rule{1.15in}{0.01in}} & \multicolumn{3}{c}{\rule{2.5in}{0.01in}}  \\
 & ($10^{9}$ M$_{\odot}$) & (M$_{\odot}$ yr$^{-1}$) & Sources & LMXBs & HMXBs & GCs & $\geq10$ Myrs & $< 10$ Myrs \\ [0.5ex] 
\hline 

NGC 0628 & 2.64 & 0.24 & 33  & 6 & 7 & 3 (50\%) & 1 (14.3\%) & 1 (14.3\%) \\ 
NGC 3351 & 7.07 & 0.50 & 30 & 9 & 8 & 3 (33.3\%) & 0$^{+2}$ (0$^{+25}$\%) & $3^{+2}$ (37.5$^{+25}$\%) \\
NGC 3627 & 19.4 & 1.81 & 48  & 15 & 14 & 3 (20\%) & 1 (7.1\%) & 1 (7.1\%) \\
NGC 4321 & 14.3 & 1.65 &  45  & 12 & 15 & 1 (8.3\%) & 0 & 0 \\ 
\textit{NGC 4569} & \textit{26.4} & \textit{1.02} & \textit{17} & \textit{10} & \textit{4} & \textit{0} & \textit{0} & \textit{0} \\ 
NGC 4826 & 19.2 & 0.39 & 20 & 15 & 4 & 3 (20\%) & 0 & 0 \\ [2ex]
& & \textbf{Total*} & 176 & 67 & 52 & 13 (19.4\%) & 2$^{+2}$ (3.8$^{+3.8}$\%)& 5$^{+2}$ (9.6$^{+3.8}$\%) \\ 
& & $\rm^{(excl.~NGC4569)}$ & & & & & & \\
\hline 
\end{tabular}
\tablecomments{The number of LMXBs and HMXBs in each galaxy are estimated using the global X-ray luminosity function fits from \citet{lehmer19}, which are a function of the total stellar mass and star formation rate enclosed within the region of interest | that is, the area within the \ks-band ellipses and covered by PHANGS. The observed cluster counts here reflect the number of XRB-hosting clusters in the region of interest, with the error indicating the most that the population could increase if the true host cluster of multiply-associated XRBs were determined, if applicable. The estimated number LMXBs and HMXBs is used to calculate the percent (in parentheses) of XRBs that appear in GCs and younger clusters, respectively. The percentages for clusters younger than 400 Myr may be inflated due to the non-negligible likelihood of a chance superposition of an XRB and young cluster in densely population regions (e.g. spiral arms). All numbers represent sources above the completeness limit of each galaxy.\\
\text{*}The total values exclude XRBs and clusters from NGC 4569, which may be significantly affected by the high X-ray completeness limit of observations.}
\end{table}

\subsection{Cluster XRBs vs. field XRBs\label{subsec:cluster_v_field}}

Studies of elliptical galaxies have shown that roughly $25-70\%$ of LMXBs are found within GCs, while the rest are found in the field \citep[e.g.][]{sarazin00,angelini01,kundu02,kundu03,maccarone03,fabbiano10,luan18}. 
A growing body of work identifying LMXBs in late-type galaxies suggests that the ratio of GC LMXBs to field LMXBs may be smaller in spiral galaxies: 
an analysis of XRBs in M101 revealed only $2\%$ of LMXBs are found in GCs \citep{chandar20}, while $12\%$ of LMXBs in M81 are found in GCs \citep{hunt22}. On the other hand, around $10\%$ of XRBs appear in clusters younger than 400 Myr in the handful of late-type galaxies studied so far \citep{chandar20, palmore22}, with a fraction of at least $25\%$ observed in NGC 4449 and the Antennae galaxies \citep{mulia19}. Tentatively, this may suggest the fraction of HMXBs in young clusters is higher than the fraction of LMXBs in GCs in late-type galaxies. 

While we do not directly classify the XRBs in this study on the basis of their donor mass, we can estimate the relative number of LMXBs and HMXBs in the target galaxies using the best-fit X-ray luminosity function of L19. Table \ref{tab:xrb_cluster_stats} shows the total number of X-ray sources within the region over which the \ks-band isophotal ellipses intersects the PHANGS region, the number of XRBs above the $90\%$ X-ray completeness limits in that region, and the estimated number of LMXBs and HMXBs. The XRB estimates take into account possible CXB contamination, so the sum of expected LMXBs and HMXBs is always lower than the total number of X-ray sources. We compare the expected number of LMXBs and HMXBs to the number that we observe within GCs and clusters younger than~$\sim 400$~Myr over the same region and above the $90\%$ completeness limit of their respective galaxies. We also calculate the percentages (in parentheses) of LMXBs found in GCs and HMXBs found in non-GCs. 

There are a few conclusions that may be drawn from the results presented in Table \ref{tab:xrb_cluster_stats}. First, the fraction of LMXBs within GCs is on the low end of the estimates from elliptical galaxies; while in early-type galaxies anywhere between $25-70\%$ of LMXBs are found in GCs, the number is between $\sim8-50\%$ within our sample of late-type galaxies, or 19.4\% over all galaxies, excluding NGC4569. This is tentatively consistent with previous studies of cluster XRBs in spiral galaxies \citep{chandar20,hunt22}, but spans a broader range of values. 

One possible cause for this apparent suppression of GC-LMXBs could be an inherent difference in the metallicity distribution of GCs in certain late-type galaxies, since the presence of XRBs in GCs is correlated with higher metallicities \citep[e.g.][]{kundu08,luan18}. The (often bimodal) metallicity distribution of GCs is thought to arise from galaxy mergers and/or GC disruption, the former of which is unlikely in spiral galaxies \citep{muratov10,pfeffer23}. Another explanation may be related to the difference in star formation rate. Since new star formation is essentially negligible in early-type galaxies, the growth of the XRB field population may slow significantly | particularly because intermediate-mass donor stars are at least a factor of 5 times more likely to form an XRB than low-mass stars in the disk of a galaxy \citep{pfahl03}, and elliptical galaxies are dominated exclusively by low-mass stars. Meanwhile, LMXBs can continue to efficiently form within GCs through dynamical interactions. This `stagnation' effect may be mitigated if GCs are responsible for seeding a significant fraction of the field LMXB population \citep[e.g.][]{lehmer20}. The difference in field-to-cluster LMXB ratios between early-type and late-type galaxies may provide further evidence that ejection from GCs has limited impact on the total LMXB population \citep[see, e.g.,][]{piro02,kundu07,peacock16,kremer18}. 

The second conclusion that may be drawn from Table \ref{tab:xrb_cluster_stats} is that the fraction of (presumably high-mass) XRBs that appear in non-GCs is almost consistently smaller than the fraction of LMXBs that appear in GCs across nearly all galaxies (except NGC 3351). These fractions do not appear directly correlated to the total \mstar\ or SFR of the host galaxy. This is inconsistent with results from M101 | for which $\sim2\%$ of LMXBs are in GCs and $\sim5\%$ of HMXBs are in very young clusters | but \textit{does} match findings from  M83 \citep[$\sim13\%$ vs. $\sim3\%$;][]{hunt21} and M81 \citep[$\sim12\%$ vs. $\sim2\%$;][]{hunt22}. Furthermore, while we do not measure metallicity directly in our cluster sample, it is interesting to note that there is no indication of a color (ergo metallicity) bias among GC LMXBs in our sample of late-type galaxies. These results are, of course, contingent on our population of clusters that host singly-associated XRBs being positively identified. As there is a non-negligible probability of an XRB within an active star-forming region being associated with an unrelated cluster (see \S\ref{sec:xclusters}), our results must be broached with caution, though previous studies must also suffer from this issue. Ultimately, there appear to be variations between galaxies that requires more exploration to fully understand.

\section{Summary}\label{sec:Summary}
This study represents one of the largest systematic investigation of XRBs in both GCs and young clusters in late-type galaxies to date. We analyze the properties of clusters that host XRBs, broken into three distinct populations: ancient GCs, intermediate-age clusters between $\sim10$ and $400$ Myr old, and very young clusters younger than 10 Myr. Our analysis of the clusters in this sample of late-type galaxies yields a number of interesting results: 

\begin{itemize}
    \item While we confirm that the chance of an XRB being wrongly associated with an unrelated, nearby GC is low (in agreement with the simulations from \citealt{hunt22}), we find there is a non-negligible probability of a chance superposition between an XRB and an unrelated cluster in densely populated, actively star-forming regions such as the spiral arms, which may inflate the statistics for clusters younger than $\sim400$ Myr. Therefore, the following results for XRBs with non-GC counterparts should be approached with caution.
    
    \item Clusters that host X-ray sources tend to be brighter than the general population for both the very young clusters and GCs. This may not necessarily be the case for intermediate-age clusters. 
    
    \item GCs in these late-type galaxies do not appear to prefer redder clusters, contrary to what has been repeatedly found for elliptical galaxies. On the other hand, there is marginal evidence that XRBs preferentially form in bluer intermediate-age clusters, though better statistics are required.
    
    \item As found in previous studies, X-ray emitting GCs tend to be more massive, denser, and have smaller effective radii than the general GC population. Among young cluster populations, there is a higher fraction of very young clusters that host XRBs in denser, more massive populations.
    
    \item There may be a strong positive correlation between cluster mass and X-ray luminosity in both intermediate-age X-ray emitting clusters and the cumulative non-GC cluster population, though a greater sample of XRBs with known cluster hosts is needed. This \textit{could} indicate the presence of multiple XRBs, though these clusters are not more X-ray luminous than the general field XRB population. On the other hand, there is a statistically significant \mstar-\lx\ \textit{anticorrelation} for GCs.
    
    \item The percentage of GCs that host XRBs in late-type galaxies varies widely between galaxies ($0.5-7\%$). This is compared to the $4-10\%$ seen in early-type galaxies. On the other hand, these percentages are $0.2-2.5\%$ for intermediate-age clusters and $0.5-1.3\%$ for young clusters, depending on the true parent clusters. 
    
    \item Whereas up to $70\%$ of LMXBs in early-type galaxies are found within their parent GCs, we find that a potentially smaller percentage ($\sim8-50\%$) are found in GCs in late-type galaxies, in agreement with statistics from previous late-type galaxy studies. 
\end{itemize}

Throughout our analysis, we note that our ability to gain insight into the population of XRBs within intermediate-age clusters is hindered by the small number of such sources in our sample. Whether this is an inherent property of cluster XRBs or a matter of insufficient data can only be addressed by a larger study of XRBs in nearby, late-type galaxies. Such a study would benefit from deeper X-ray observations (including a revisit of NGC 4569), optical observations covering a larger sky area, X-ray and optical spectroscopy, and metallicity measurements of compact star clusters. This type of data would give us a clearer picture of these sources by allowing us to identify black hole vs. neutron star XRBs within and around clusters, constrain the cluster-to-field ratios, and study the cluster metallicity dependence of XRBs in late-type galaxies.  

\begin{acknowledgments}
Q.H. is partially funded by a Rackham Merit Fellowship,
awarded by the University of Michigan Rackham Graduate School. Q.H. thanks the LSSTC Data Science Fellowship Program, which is funded by LSSTC, NSF Cybertraining Grant no. 1829740, the Brinson Foundation, and the Moore Foundation; her participation in the program has benefited this work. 

The Digitized Sky Surveys were produced at the Space Telescope Science Institute under U.S. Government grant NAG W-2166. The images of these surveys are based on photographic data obtained using the Oschin Schmidt Telescope on Palomar Mountain and the UK Schmidt Telescope. The plates were processed into the present compressed digital form with the permission of these institutions.
\end{acknowledgments}

%

\vspace{5mm}
\facilities{CXO, HST(STIS), ALMA, VLT}


\software{Astropy \citep{astropy}, Cigale \citep{boquien19}, BAOlab \citep{larsen99}, SciPy \citep{scipy}
          }





\bibliography{Hunt_2022}{}
\bibliographystyle{aasjournal}



\end{document}